\documentclass[twocolumn,showpacs,preprintnumbers,amsmath,amssymb,pre,longbibliography]{revtex4-1}

\usepackage{color}    
\usepackage{graphicx}
\usepackage{braket} 
\usepackage{dcolumn}
\usepackage{bm}
\usepackage{subfigure}
\usepackage{amssymb}
\usepackage{multirow}
\usepackage{natbib}
\usepackage{amsmath}
\usepackage{mathtools}
\graphicspath{{plots/}}
\usepackage[english]{babel}

\newcommand{\db}[2][]{\text{d}^{#1}#2}

\DeclarePairedDelimiter\Avr{\langle}{\rangle}%
\newcommand*{\al}{\alpha}
\newcommand*{\be}{\beta}

\newcommand{\sgn}{\text{Sgn}}
\newcommand{\upspin}{\uparrow}
\newcommand{\downspin}{\downarrow}

\renewcommand{\vec}[1]{\mathbf{#1}}

\DeclareMathOperator\erf{erf}
\DeclarePairedDelimiter\abs{\lvert}{\rvert}%

\begin{document}

% \keywords{plasma crystals, streaming effects, Yukawa balls, string formation, dynamical screening}

%\title[The ion potential in warm dense matter]{The ion potential in warm dense matter: wake effects due to streaming degenerate electrons}
%\title{Quantum plasma theory--quo vaids?}
\title{The equation of state of partially ionized hydrogen and deuterium plasma revisited}

\author{A.V. Filinov}
\email{filinov@theo-physik.uni-kiel.de}
\affiliation{Institut für Theoretische Physik und Astrophysik, Christian-Albrechts-Universität zu Kiel, Leibnizstra{\ss}e 15, 24098 Kiel, Germany}

\author{M. Bonitz}\email{bonitz@physik.uni-kiel.de}
\affiliation{Institut für Theoretische Physik und Astrophysik, Christian-Albrechts-Universität zu Kiel, Leibnizstra{\ss}e 15, 24098 Kiel, Germany}

%\author{ A.~Filinov$^1$,  and M.~Bonitz$^1$
%}
%\affiliation{
% $^1$Institut f\"ur Theoretische Physik und Astrophysik, Christian-Albrechts-Universit\"at zu Kiel,
% Leibnizstra{\ss}e 15, 24098 Kiel, Germany}
%\affiliation{$^2$IVTAN, Russia}

%---------------
\begin{abstract}
We present novel first-principle fermionic path integral Monte Carlo (PIMC) simulation results 
%without fixed nodes 
for a dense partially ionized hydrogen (deuterium) 
plasma, for temperatures in the range $15,000$K $\leq T \leq 400,000$K and densities $7 \cdot 10^{-7}$g/cm$^{3}\leq \rho_H \leq 0.085$ g/cm$^{3}$ ($1.4 \cdot 10^{-6}$g/cm$^{3}\leq \rho_D \leq 0.17$ g/cm$^{3}$),
corresponding to $100\geq r_s\geq 2$, where $r_s=\bar r/a_B$ is the ratio of the mean interparticle distance to the Bohr radius. These simulations are based on the fermionic propagator PIMC (FP-PIMC) approach in the grand canonical ensemble [A. Filinov \textit{et al.}, Contrib. Plasma Phys. \textbf{61}, e202100112 (2021)] and fully account for correlation and quantum degeneracy and spin effects. For the application to hydrogen and deuterium, we develop a combination of the fourth-order factorization and the pair product ansatz for the density matrix.
Moreover, we avoid the fixed node approximation that may lead to uncontrolled errors in restricted PIMC (RPIMC). Our results allow us to critically re-evaluate the accuracy of  the RPIMC simulations for hydrogen by Hu \textit{et al.} [Phys. Rev. B \textbf{84}, 224109 (2011)] and of various chemical models. The deviations are generally found to be small, but for the lowest temperature, $T=15,640$~K they reach several percent. We present detailed tables with our first principles results for the pressure and energy isotherms.
\end{abstract}

%52.25.Dg Plasma kinetic equations
%52.27.Aj Single-component, electron-positive-ion plasmas
%52.27.Gr non-ideal plasmas
%52.65.Yy Molecular dynamics methods
%Plasma simulation, 52.65.-y
\maketitle

 \section{Introduction}\label{s:intro}
Warm dense matter (WDM) is a rapidly growing research field on the boarder of plasma physics and condensed matter physics, e.g.~\cite{graziani-book,Fortov2016, moldabekov_pre_18, dornheim_physrep_18}. Examples include astrophysical objects such as the plasma-like matter in brown and white dwarf stars \cite{saumon_the_role_1992, chabrier_quantum_1993,chabrier_cooling_2000}, giant planets, e.g. \cite{schlanges_cpp_95,bezkrovny_pre_4, vorberger_hydrogen-helium_2007, militzer_massive_2008, redmer_icarus_11,nettelmann_saturn_2013}  and the outer crust of neutron stars \cite{Haensel,daligault_electronion_2009}. In the laboratory, WDM is being routinely produced via laser  compression \cite{nora_gigabar_2015} or with Z-pinches \cite{matzen_pulsed-power-driven_2005,knudson_direct_2015} and, in the near future, also via ion beam compression \cite{hoffmann_cpp_18,tahir_cpp19}.  
A particularly exciting application is inertial confinement fusion (ICF) \cite{moses_national_2009,hurricane_inertially_2016} 
 where recently important breakthroughs have been achieved \cite{PhysRevLett.129.075001, PhysRevE.106.025202, PhysRevLett.131.015102}.

In warm dense matter research and ICF, in particular, an accurate description of hydrogen (and deuterium) plays a central role. Hydrogen -- the most abundant and, at the same time, the simplest element in the universe -- has been in the focus of research for many decades. Its properties have been investigated, among others, in many compression experiments with high intensity lasers, at  x-ray free electron laser facilities and at the National Ignition Facility (NIF) at Livermore National Laboratory. 
Among questions of particular interest are the predicted metal--insulator transition, the hypothetical plasma phase transition and the recently predicted roton feature  \cite{dornheim_prl_18,dornheim_comphys_22,hamann_prr_23}.
Aside from these questions, also basic properties such as the equation of state (EOS), compressibility, optical properties and conductivity are of prime importance for experiments involving hydrogen. However, the accuracy of existing experimental data is still not fully known. The situation is expected to change with upcoming high accuracy experiments including the colliding planar shocks platform at the NIF \cite{nif-eos_23} and the FAIR facility at GSI Darmstadt \cite{tahir_cpp19}.

All this poses particular challenges to theory which comprises a broad arsenal of models and simulation approaches. Even after four decades of research there persist significant deviations among the results of different models, and there remains a surprisingly large scatter of data, even for relatively simple quantities, such as the equation of state. 
The reason is that the relevant thermodynamic and transport properties of warm dense hydrogen are influenced, among others, by electronic quantum effects, moderate to strong Coulomb correlations, the formation of atoms and molecules, lowering of the ionization and dissociation potentials, and the interaction between charged and neutral species. 
Available data include (semi-)analytical results within the chemical picture, such as the Pade formulas of Ebeling \textit{et al.} \cite{Ebeling_1985,trigger_jetp_03}, the 
    Saumon, Chabrier van Horn model \cite{chabrier_quantum_1993}, the model of Schlanges \textit{et al.}~\cite{schlanges_cpp_95}, and the fluid-variational theory of Redmer \textit{et al.}~\cite{Juranek2002}, 
semi-classical molecular dynamics with quantum potentials (SC-MD), e.g.~\cite{filinov_pre04}, electronic force fields \cite{goddard_PhysRevLett.99.185003,dai_PhysRevLett.122.015001} and various variants of quantum MD, e.g. \cite{filinov_prb_2, knaup_cpc_02,sjostrom_prl_14,dai_PhysRevLett.104.245001,Kang_2018}.
A recent breakthrough occurred with the application of Kohn-Sham density functional theory (DFT) and Born-Oppenheimer MD simulations because they, for the first time, enabled the selfconsistent simulation of realistic warm dense matter, that includes both, plasma and condensed matter phases, e.g. \cite{collins_pre_95,plagemann_njp_2012,witte_prl_17}. Further developments include  orbital-free DFT methods (OF-DFT), e.g. \cite{KARASIEV20143240,dai_doi:10.1016/j.mre.2017.09.001} and  time-dependent DFT (TD-DFT), e.g.  \cite{baczewski_prl_16}. 

However, all these methods involve severe approximations -- that are related e.g. to the concept of the chemical picture, the semi-classical approximation in MD or the choice of the exchange-correlation functional in DFT -- that give rise to systematic errors that are difficult to quantify and that strongly limit the predictive power of the methods. Therefore, a special role is played by first principle computer simulations that -- at least in principle -- are free of systematic errors and, therefore, may serve as benchmarks and reference for other models. For the thermodynamic quantities of warm dense hydrogen such a role is being played by path integral Monte Carlo (PIMC) simulations, pioneered by D.M.~Ceperley and B.~Militzer \cite{militzer_path_2000,Militzer2001}, as well as V.S.~Filinov and co-workers, e.g. \cite{filinov_jetpl_00,filinov_pla_00,filinov_ppcf_01,filinov_cpp_01}. However, PIMC simulations for warm dense hydrogen are severely hampered by the necessity to accurately treat the Fermi statistics of the electrons. This is known as the fermion sign problem (FSP) that was shown to be N-P hard 
\cite{Troyer2005,Dornheim_jpa_2021}, i.e. simulations suffer an exponential loss of accuracy with increasing quantum degeneracy of the electrons. For this reason, the computationally expensive direct fermionic PIMC simulations of V.S.~Filinov \textit{et al.}, that could use only a limited number of high-temperature factors on the order of $20\dots 50$, became increasingly inaccurate for values of the electron degeneracy parameter $\chi \gtrsim 1$, cf. Eq.~\eqref{eq:theta-chi}. An alternative strategy was developed by Ceperley: eliminate the FSP via the fixed node approximation \cite{Ceperley1991,sign_cite}. This leads to the restricted PIMC (RPIMC) method, and the RPIMC results for hydrogen, including new data \cite{hu_militzer_PhysRevLett.104.235003,Milizer2011}, still today remain the most reliable reference. The importance of the RPIMC data has grown even more because they have been used by Militzer and co-workers as primary input (in combination with DFT) for extended thermodynamic data tables for hydrogen and a broad set of more complex materials \cite{Driver_2015}.

The magnitude of the systematic error in RPIMC that is introduced by the fixed node approximation is commonly believed to be small. This is, in part, based on the good accuracy of the associated zero-temperature approach (diffusion Monte Carlo), as confirmed by available full CI ground state data. However, the situation is different at finite temperature. While the fixed node approximation in RPIMC has been criticized \cite{filinov_jpa_2001,filinov_ht_2014}, and occasionally improvements beyond free-particle nodes have been tested \cite{Militzer_2000,Militzer_2021}, so far, no alternative first-principle results have been available that allow one to assess the accuracy and applicability range of the method at finite temperatures.
This has changed only recently, when configuration PIMC (CPIMC) simulations have been introduced by Bonitz and co-workers~\cite{schoof_cpp11,schoof_cpp15}. These are first principle finite temperature PIMC simulations in Fock space that have no sign problem at strong degeneracy and are thus complementary to standard coordinate space PIMC. When applied to the model system of the warm dense uniform electron gas (UEG), the combination of CPIMC with permutation blocking PIMC -- coordinate space PIMC that uses fourth-order propagators and determinant sampling -- and was introduced by Dornheim (PB-PIMC \cite{dornheim_njp15}) made it possible to produce benchmark thermodynamic data with a relative error below $1\%$ \cite{dornheim_prb16, groth_prb16,
dornheim_physrep_18}. These results were used, among others, to develop finite temperature exchange-correlation functionals for DFT, such as the finite-temperature LDA functional GDSMB \cite{groth_prl17}. Moreover, these UEG results provided, for the first time, the opportunity to analyze the accuracy of RPIMC simulations in the WDM range. In fact comparison to RPIMC simulations for the UEG of Brown \textit{et al.}~\cite{Brown_2014}  revealed unexpectedly large errors of the latter exceeding $10\%$ around $r_s=1$ \cite{schoof_prl15}, for more details, see Refs.~\cite{dornheim_pop17,dornheim_physrep_18,filinov_cpp_21}.

It is, therefore, of high interest to extend the first-principle direct PIMC simulations to warm dense hydrogen, applying the novel advanced methods that were successfully applied to the UEG before. This will also allow one to benchmark the aforementioned earlier hydrogen simulations and RPIMC, in particular. This is the goal of the present paper. To this end we apply the fermionic propagator PIMC (FP-PIMC) approach -- a grand canonical extension of the PB-PIMC method that was recently successfully applied to the thermodynamic properties of the UEG in Ref.~\cite{filinov_cpp_21} and to the static density response function and the dynamic structure factor of the UEG \cite{filinovDSF_2023} and helium 3 \cite{filinovHe3_2023}.

Here we extend the FP-PIMC method to a two-component electron-proton plasma. The approach is formulated in coordinate space and, therefore, the fermion sign problem restricts our simulations to moderate densities, $r_s\gtrsim 2$. We present extensive new data for a broad density and temperature range, $r_s \in [2; 100]$ and $T\in [15,000; 400,000]$. Our results include the equation of state, various energy contributions, and pair distributions. Further, we extract approximate data for the degree of ionization and the fraction of molecules. The comparison with earlier results reveals significant inaccuracies of the latter, in particular for low temperatures and/or low densities.

 This paper is organized as follows: in Sec.~\ref{s:hydrogen-overview} we recall the main parameters and  give a brief overview on selected theoretical data for warm dense hydrogen which will be used for comparison to our results. In Sec.~\ref{s:overview-our-pimc} our FP-PIMC approach is presented and its extension to hydrogen is explained. 
Section~\ref{s:results} presents our numerical results. Detailed data tables for the equation of state and total energy are presented in the appendix.

%------------
\section{Warm dense hydrogen and deuterium} \label{s:hydrogen-overview}
\subsection{Relevant parameters}\label{s:qp_parameters}
Let us recall the basic parameters of warm dense hydrogen \cite{bonitz_qkt,bonitz_pop_19}: 
\begin{itemize}
    \item The first are the quantum degeneracy parameters, $\theta$ and $\chi$, which involve the Fermi energy $E_F$ and the thermal DeBroglie wave length $\lambda$, respectively:
\begin{align}
\theta &= \frac{k_B T}{E_{F}}\,, \quad \chi=n\lambda^3\,,\label{eq:theta-chi}\\
\lambda & = \frac{h}{\sqrt{2\pi m k_B T}}\,,
\label{eq:lambda}
\\
    E_{F} &= \frac{\hbar^2}{2m}(3\pi^2 n)^{2/3}\,,
    \label{eq:ef-def}
\end{align}
where $n$ is the density, $T$ the temperature, and $m$ the particle mass. The proton degeneracy parameter $\chi_p$ is a factor $(m_e/m_p)^{3/2}$ smaller than the one of the electrons and is negligible in the parameter range studied in this paper.
%    \item 
\item The second parameter is the classical coupling parameter of protons, 
\begin{align}
\Gamma_p= \frac{e^2}{a k_BT}\,,\quad    a^3=\frac{4}{3\pi n}\,,
\end{align}
where $a$ is the mean inter-particle distance.
\item Finally, the quantum coupling parameter (Brueckner parameter) of electrons in the low-temperature limit is,
    \begin{equation}
    r_s = \frac{a}{a_B}, \quad a_B = \frac{\hbar^2}{m_r e^2}, \quad m_r= \frac{m_e m_p}{m_e+m_p},  
    \label{eq:rs-def}
    \end{equation}
    where $a_B \approx 0.529\AA$ is the Bohr radius, and $m_r$ is the reduced mass which, for hydrogen and deuterium, is $m_r \approx m_e$. 
\end{itemize}

Note that the degeneracy parameters refer to an ideal plasma and give only a rough picture of the physical situation in WDM. At moderate to strong coupling, the spatial extension of electrons may be strongly modified and the physical degeneracy parameter may  differ substantially from $\theta$ and $\chi$. Similarly, the  presence of bound states significantly alters the physical degeneracy parameters.
For a discussion of this effect, see Ref.~\cite{filinov_jpa03}. On the other hand, bound states also significantly affect the coupling strength in the plasma because they lead to a reduction of the number of free particles that interact via the long range Coulomb potential. 

\subsection{Selected theoretical reference results for warm dense hydrogen} \label{ss:hydrogen-results}

Here we give a brief overview on existing theoretical and simulation results for dense hydrogen and deuterium, see also the discussion in Sec.~\ref{s:intro}. Here we concentrate on those that will be considered for comparison with our simulation results in Sec.~\ref{s:results}.
For further details on the different methods, the reader is referred to various text books, e.g.~\cite{afilinov_rinton,Fortov2016,graziani-book}.

\begin{enumerate}
    \item 
by \textbf{CP2019} we denote the hydrogen EOS by Chabrier {\it et al.}~\cite{Chabrier_2019} that combines simulations from three relevant physical domains, for temperatures below $T\leq  100, 000$K: 
\begin{description}
\item[i.] $\rho \leq 0.05$ g/cm$^{3}$ ($r_s \geq 3.8$): the theory of Saumon, Chabrier and van Horn~\cite{saumon_horn_1995} in the low-density, low-temperature molecular/atomic phase, 
\item[ii.] $\rho > 10$ g/cm$^{3}$ ($r_s < 0.65$): the EOS of Chabrier and Potekhin~\cite{Chabrier_1998} in the fully ionized plasma (the high-density and  high-temperature phase),  
\item[iii.] $0.3 \leq \rho \leq  5$ g/cm$^{3}$ ($0.8 \leq r_s \leq 2$) ab initio quantum molecular dynamics calculations~\cite{Caillabet_2011} at intermediate density and temperature, dominated by pressure dissociation and ionization processes.    
\item[iv.] For the missing density interval ($2 \leq r_s \leq 3.8$) spline interpolation of the main thermodynamic quantities is performed that ensures continuity of the quantities and their first two derivatives.
\end{description}
\item by \textbf{FVT} we denote fluid variational theory of  Juranek \textit{et al.} \cite{Juranek2002}. There hydrogen is treated as fluid mixture of atoms and molecules, including deviations from linear mixing, and their concentrations are obtained by self-consistent solutions of non-ideal Saha equations. That reference states reasonable accuracy for $\rho \lesssim 0.6$ g cm$^{-3}$.
%\item by \textbf{H-REOS.3} we denote the wide-range Rostock EOS of Ref.~\cite{Becker_2014}. It combines a variety of data sets: 
%\begin{description}
%    \item[i.] $T\lesssim 10^4$K: at low density ($\rho \lesssim 0.1$g cm$^{-3}$), fluid variational theory (FVT) and a modification that includes ionization (FVT+ \cite{Holst2007}) is used, and for higher densities DFT-MD \textit{ab initio} data.
%    \item[ii.] $T\gtrsim 10^4$K: a combination of the Saumon-Chabrier-van Horn EOS \cite{saumon_horn_1995}, DFT-MD and a  RPIMC-based interpolation is used.
%    \item[iii.] $T\gtrsim 10^5$K: the Chabrier-Potekhin model \cite{Chabrier_1998} is used (also for densities exceeding $\rho=10^2$ gcm$^{-3}$).
%\end{description}
\item by \textbf{WREOS} we denote the wide-range EOS of Wang and Zhang, Ref.~\cite{Wang2013}. They combine \textit{ab initio} Kohn-Sham DFT-MD, for $\theta < 1$ with orbital-free DFT simulations, for $\theta > 1$.
\item  by \textbf{HXCF} we denote the equation of state table of deuterium of Mihaylov \textit{et al.}, Ref.~\cite{Karasiev_2021}. It aims at a universal DFT-based treatment for all temperatures and densities including high-order exchange-correlation functionals and also path integral MD (PIMD) data. The data points for $r_s=17.53$ (deuterium density $\rho_D=0.001$g cm$^{-3}$) that are included in the figures below have been obtained by Kohn-Sham MD (based on PIMD).
\item by \textbf{RPIMC} we denote results from restricted (fixed nodes) PIMC simulations and RPIMC-DFT combinations, see also Sec.~\ref{s:intro}, for a discussion. In Ref.~\cite{Militzer_2021} a first-principles equation of state database was provided which will be used for comparison throughout this work. While these tables are given for many materials and contain combinations of RPIMC data with DFT-LDA simulations, for hydrogen only RPIMC data are involved~\cite{Militzer_2021}.

\end{enumerate}
The above selection of the different hydrogen equations of state is by no means representative. The goal is to compare our new PIMC data with recent results that are frequently used in astrophysics or warm dense matter research and for which reference data for the same temperatures have been published so interpolation can be avoided. 

In our opinion, among the data above, the RPIMC-based ones are the most reliable because they do not involve any sources of systematic errors -- except for the choice of the nodes in the fixed node approximation. On the other hand, many other equations of state involve RPIMC data, in one way or the other. For these reasons, the comparison with RPIMC is in the focus of the analysis of our new simulation data. Since our FP-PIMC simulations do not involve any  approximation, such as fixed nodes, we expect them to be more reliable as RPIMC and, in case of deviations between the two, the origin should be in the nodes of RPIMC. The main source of error in FP-PIMC is of statistical nature, for these reasons we devote special attention to verify convergence of our results.

%\subsection{DFT simulations for dense hydrogen}\label{ss:dft}
%\textcolor{red}{MB: check DFT data by Militzer and finite-T data by Kushal}

%there are 2 figures in this project that can be used for comparison (they will be published in the Phys. Plasmas paper). 

\section{Overview on the present direct fermionic PIMC simulations}\label{s:overview-our-pimc}

\subsection{Fermionic propagator PIMC (FP-PIMC)}\label{ss:pb-pimc-plasma}
We use Feynman's path integral picture of quantum mechanics where particles are represented by coordinate space-imaginary time ``trajectories''. Fermions with different spin projections are denoted by the coordinate vector $\vec R^{s}_{p}=(\vec r^s_ {1,p},\ldots,\vec r^s_ {N^s,p})$ with $\{s=\uparrow,\downarrow \}$,  and the ions by the vector  $\vec R^{I}_{p}$. The ensemble of the particle trajectories in the imaginary time is denoted by the variable  $\vec X^{s(I)}=(\vec R^{s(I)}_1,\ldots, \vec R^{s(I)}_P)$, where the lower index indicates the imaginary time argument, $\tau_p= p \epsilon$ with $\epsilon=\beta/ P$ and $0\leq p\leq P$. Here, $P\ge 1$ denotes the number of high-temperature factors (``time slices'') in the path integral formalism, see below.

To evaluate the fermionic partition function, the sum over $N_s!$ permutations is performed explicitly,  which contains sign-alternating terms, where the sign $\sgn(\sigma_s)=\pm 1$ depends on the parity of the permutation in each of the two spin-subspaces: 
\begin{align}
Z_F=&(N^{\uparrow}! N^{\downarrow}!)^{-1}  \sum\limits_{p \in \wp} \int \db \vec X^{\uparrow} \db \vec X^{\downarrow} \db \vec X^{I} \prod
\limits_{p=1}^P \sgn(\sigma_\uparrow) \sgn(\sigma_\downarrow) \nonumber \\
&\times \abs{\langle \vec R_{p-1}|e^{-\epsilon \hat H} |\hat \pi_{\sigma_\uparrow} \hat \pi_{\sigma_\downarrow} \vec R_{p}\rangle}\,,
\label{Zf}
\end{align}
where we introduced the notation $\vec R_{p}=(\vec R^{\uparrow}_p,\vec R^{\downarrow}_p, \vec R^I_p)$, and the sum over $p \in \wp$ runs over different permutation classes. 
In this representation, called the {direct} path-integral Monte Carlo (DPIMC) method, physical expectation values are evaluated with the statistical error
\begin{eqnarray}
\Avr{A}=\frac{\left\langle\hat A \cdot \prod\limits^s\sgn(\sigma_s)\right\rangle}{\Avr{s}} \pm \delta A \label{AvrA},\quad \delta A \sim 1/\Avr{s}\,,\label{meanA}
\end{eqnarray}
which scales inversely proportional to the average sign
\begin{eqnarray}
\Avr{s}=\left\langle\prod\limits^s \sgn(\sigma)\right\rangle=\frac{Z_F}{Z_B}=e^{-\beta N(f^F-f^B)} \,.\label{AvrS}
\end{eqnarray}
The sign decays exponentially with particle number $N$, inverse temperature $\beta$ and the free energy difference of bosons and fermions. 
%A direct illustration of the inefficiency of this approach can be recovered in the simulations of ideal or interacting Fermi gas (jellium model) ~[]. For this reason, the DPIMC can not resolve a fundamental problem of  ab initio treatment of  many-body fermionic systems.

To recover the effect of the exchange-correction hole as observed by the interaction of two spin-like fermions, one important improvement to the DPIMC sampling~(\ref{Zf}) is necessary. This physical effect can be reproduced in numerical simulations by the use of anti-symmetric propagators (determinants) already on the level of stochastic importance sampling of particle trajectories.
This requires to introduce a modified partition function where the summation over different permutation classes is performed analytically in the kinetic energy part of the density matrix (DM), and results in the Slater determinant constructed from the free-particle propagators  
\begin{eqnarray}
&&Z_F=\frac{1}{N^{\uparrow}! N^{\downarrow}!} \int \db \vec X^{\uparrow} \db \vec X^{\downarrow} \db \vec X^{I} \prod\limits_{p=1}^P \sgn_p \, e^{-S^A_{p}}\label{ZF2} \\ 
&&\sgn_p=\sgn \det \mathbb{M}^{\uparrow}_{p-1,p} \cdot \sgn \det \mathbb{M}^{\downarrow}_{p-1,p}, \\
&& e^{-S^A_{p}}= e^{-\epsilon U(\vec R_{p-1},\vec R_{p})} e^{W_{\text{ex}}(\vec R_{p-1},\vec R_{p})},\\
&& W_{\text{ex}}=\ln \abs{\det \mathbb M^{\uparrow}_{p-1,p}}+\ln \abs{\det \mathbb M^{\downarrow}_{p-1,p}},
\end{eqnarray}
where the off-diagonal ``action'', $S^A_{p}$, depends on the interaction term $U$ between charged particles on two successive time slices ($p-1,p$), while $W_{\text{ex}}$ accounts for exchange effects due to Fermi statistics via the Slater determinants.

The fermionic (anti-symmetric) free-particle (FFP-) propagator between two adjacent time slices is given by 
 \begin{align}
 D_{p-1,p}^s=\sum\limits_{\sigma_s} \left\langle \vec R^s_{p-1} \left|e^{- \epsilon \hat K} \right| \hat \pi_{\sigma_s} \vec R^s_{p} \right\rangle=\frac{\det \mathbb M^s_{p-1,p}}{\lambda^{DN^s}_\epsilon},
%\nonumber
\label{M1}
\end{align}  
where the spin components are denoted by $s=\uparrow,\downarrow$, and  $\vec M^s$ is the $N^s \times N^s$ anti-symmetric diffusion matrix
\begin{eqnarray}
&&\mathbb M^s=||m_{kl}(p-1,p)||, \quad k,l=1,\ldots N^s\,,\\
&&m_{kl}(p-1,p)=\exp\left(-\frac{\pi}{\lambda^2_{\epsilon}}[\vec r^s_{l,p}-\vec r^s_{k,p-1}]^2\right)\,.   \label{M2}
\end{eqnarray} 

This representation has several advantages. First, the resulting density matrix is anti-symmetric with respect to any pair exchange of same spin fermions. Second, the probability of micro-states is now proportional to the value of the Slater determinants, and the degeneracy of the latter, at small particle separations, correctly recovers the Pauli-blocking effect. Third, the FFP-propagators partially reduce the fermion sign-problem (FSP).

The change of the sign of the Slater determinants, evaluated along the imaginary time interval, $0\leq \tau_p\leq \beta$, is taken into account by the extra factors, $\sgn_p$. Combined together they define the average sign in  fermionic PIMC, 
\begin{eqnarray}
\Avr{S}=\left\langle\prod\limits_{p=1}^P\sgn \, \det \mathbb{M}^{\uparrow}_{p-1,p} \cdot \sgn \det  \mathbb{M}^{\downarrow}_{p-1,p}\right\rangle,\label{def_Sign}   
\end{eqnarray}
and characterize the efficiency of simulations in terms of the statistical error $\delta A$ in Eq.~(\ref{meanA}). The PIMC simulations become hampered by the FSP~\cite{CeperleyFermi,Troyer2005} once the statistical uncertainties are strongly enhanced due to an exponential decay of the average sign $\Avr{S(N,\beta)}$ with the particle number $N$, the inverse temperature $\beta=1/k_B T$ or the degeneracy parameter, $\theta$ (or $\chi$).
The usage of the {\it fermionic propagators}, Eq.~(\ref{M1}), permits to partially overcome the sign problem and make the simulations feasible up to $\chi \lesssim 3$.

The efficiency of the fermionic propagator approach has been demonstrated by several authors, including Takahashi and Imada~\cite{Takahashi1984}, V. Filinov and co-workers~\cite{filinov_ppcf_01,filinov_jetpl_01,bonitz_prl_5} and Lyubartsev~\cite{Lyubartsev_2005}. Chin~\cite{Chin2015} used determinant propagators  to simulate relatively large ensembles of electrons in 3D quantum dots. The PB-PIMC simulations by  Dornheim \textit{et al.} \cite{dornheim_njp15,dornheim_physrep_18} have addressed the thermodynamic proprieties of the UEG from first principles.

\subsection{Quantum pair potentials in plasma simulations}\label{ss:qpp}
Before discussing our plasma simulations, we review the concept of effective quantum potentials that are used to overcome the divergence of the attractive Coulomb potential which leads to divergencies in classical statistical thermodynamics. Taking quantum effects into account gives rise to modified pair potentials which do not exhibit these singularities anymore. Such potentials have been derived by Kelbg \cite{kelbg_ap_63,kelbg_ap_63_2,kelbg_ap_64}, Deutsch \cite{deutsch_pla_72} and others. They capture the basic quantum diffraction effects and are exact in the weak coupling limit. Furthermore, an improved version of the Kelbg potential (IKP) was derived \cite{filinov_pre04,filinov_jpa03} that extends the applicatbility range beyond the weak coupling limit, as will be discussed below. In Eq.~\eqref{Kelbg} we reproduce the IKP whereas
the original Kelbg potential follows from it by setting the parameter $\gamma_{ij}$ equal to one.
\begin{eqnarray}
 &&U(r,\lambda_{ij})=\frac{Z_q}{r} \left[1- e^{-\frac{r^2}{\lambda_{ij}^2}} +\frac{\sqrt{\pi} r}{\lambda_{ij} \gamma_{ij}}\left(1-\erf \left[\frac{\gamma_{ij} r}{\lambda_{ij}}\right]\right) \right], \nonumber \\
&&  Z_q=q_i q_j,\quad \lambda_{ij}^2=\frac{\hbar^2 \beta}{2 \mu_{ij}},\quad \frac{1}{\mu_{ij}}=\frac{1}{m_i}+\frac{1}{m_j}  \,,\label{Kelbg}
\end{eqnarray} 
and was obtained via first-order perturbation theory. Quantum effects related to the finite extension of particles on the length scale of the de-Broglie wavelength, $\lambda_{ij}$, which depends on temperature and the reduced mass, enter explicitly. 

This potential has the advantage of preserving the correct first derivative of the the exact binary Slater sum $S_{ij}(r)$ at zero interparticle distance. However, it does not include strong coupling effects, in particular it does not include bound states, which become important 
%in the case of electron-proton interaction 
at low temperatures. This  drawback has been corrected with the introduction of the {\em improved Kelbg potential} (IKP) by A. Filinov \textit{et al}.~\cite{vfilinov_jpa_03, filinov_pre04}. The  correction parameter $\gamma_{ij}$ in Eq.~(\ref{Kelbg}) is directly related to the exact Slater sum at zero distance 
\begin{eqnarray}
 \gamma_{ij}(\beta,\mu_{ij})=-\frac{\sqrt{\pi}}{\lambda_{ij}}\frac{Z_q \beta}{\ln [S_{ij}(r_{ij}=0,\beta)]}\,.
\end{eqnarray}
A detailed discussion of the accuracy of the IKP, Eq.~\eqref{Kelbg}, for all types of binary Coulomb interactions, and practical applications for a hydrogen plasma in both, PIMC and MD simulations was presented in Ref.~\cite{filinov_pre04}. 

In our recent work~\cite{bonitz_cpp_23} we performed detailed accuracy and convergence tests, {being relevant for the present plasma simulations. First, we performed FP-PIMC simulations with the IKP and compared them to simulations where the exact off-diagonal pair potential (see next Sec.~\ref{4thorderP}) for the electron-ion interaction~\cite{Ceperley1995} (defined by the exact Slater sum) was employed.
In summary, we found that, while the IKP  allows to accurately describe the electron-electron (ion-ion) correlation effects, its accuracy given by the diagonal approximation to the off-diagonal pair density matrix 
%in PIMC simulations, 
%for the electron-ion 
%interaction the accuracy of the diagonal approximation to the off-diagonal pair density matrix,
\begin{equation}
\rho(\vec{r},\vec{r}^{\prime}; \tau)\approx \frac{1}{2} \left[\rho(\vec{r},\vec{r};\tau)+\rho(\vec{r}^{\prime},\vec{r}^{\prime};\tau) \right]\,, \label{diagDM}   
\end{equation}
is not sufficient. The IKP  exhibits very slow convergence with respect to the number of high-temperature propagators $P$, as was shown in Ref.~\cite{bonitz_cpp_23}.
Additional extensive tests of the $P$-convergence in the present FP-PIMC simulations, including the dependence on the system size, will be 
summarized in Sec.~\ref{ss:convergence}.

\subsection{Combination of the fourth-order factorization scheme with the product density matrix ansatz}\label{4thorderP}
As shown in the previous discussion, while it is sufficient to use the fitted IKP for the binary interactions $ij=ee,ii$, a more accurate treatment of the attractive electron-ion interaction is required to reduce the number of factors, $P$, to a moderate value. This is crucial because the efficiency of FP-PIMC sensitively depends on the imaginary time step $\epsilon=\beta/P$. 
A larger time step (smaller $P$-value)
in the high-temperature factorization increases the average sign $\Avr{S}$ in Eq.~(\ref{def_Sign}) and extends the applicability range of fermionic simulations to higher degeneracy. To achieve this goal -- without loss of accuracy -- as was done in the UEG case~\cite{filinov_cpp_21,dornheim_njp15}, we take advantage of the fourth-order factorization scheme proposed by 
Chin \textit{et al.}~\cite{Chin2002} and Sakkos \textit{et al.}~\cite{Sakkos2009}:
\begin{eqnarray}
  &&e^{-\beta\hat H}
     =\prod\limits_{p=1}^P e^{-\epsilon(\hat K+\hat V) } \label{4thorder} \\
  &&\approx \prod\limits_{p=1}^P e^{-\epsilon \hat W_{1}} e^{-t_1 \epsilon \hat K}  e^{- \epsilon \hat W_{2}} e^{-t_1 \epsilon \hat K} e^{-\epsilon \hat W_{1}} e^{-t_0 \epsilon \hat K} + O(\epsilon^{4})\,.
    \nonumber
\end{eqnarray}
The potential and the kinetic energy contributions along the imaginary time step $\epsilon$ are divided into three parts as, $t_1\epsilon+ t_1\epsilon +t_0 \epsilon=\epsilon$, with $t_0$ being a free parameter which influences the $P$-convergence. As a result 
the higher order commutators between $\hat K$ and $\hat V$  exactly cancel, up to the order $O(\epsilon^4)$~\cite{Suzuki1985}. Thereby the effective total number of factors in Eq.~(\ref{ZF2}) becomes $3P$, which has to be taken into account in the cited $P$-values in Sec.~\ref{ss:convergence}.

The new potential energy operators introduced in (\ref{4thorder}) take the form~\cite{Chin2002}
\begin{eqnarray}
    &&\hat W_1=v_1 \hat V+u_0 a_1  \hat W_Q,\label{W1}\\
    &&\hat W_2=v_2 \hat V+u_0 a_2  \hat W_Q,\quad a_2=1-2 a_1\label{W2} \\
    &&\hat W_Q=\epsilon^2 [[\hat V,\hat K],\hat V]=\frac{\hbar^2}{m} \epsilon^2 \sum\limits_{i=1}^N \abs{\vec F_i^2}, \label{W}
\end{eqnarray}
where $\vec F_i$ is the full force acting on a particle ``i''.  
The involved coefficients are defined as
\begin{eqnarray}
&&v_1=\frac{1}{6 (1-2 t_0)^2},\; v_2=1-2 v_1,\; 2t_1=1-t_0,\\
&&u_0=\frac{1}{12} \left[ 1-\frac{1}{1-2 t_0} +\frac{1}{6(1-2 t_0)^3} \right].
\end{eqnarray}
The choice $t_0=1/6$ (as used here), in particular, corresponds to an equidistant time-step in Eq.~(\ref{4thorder}), and a symmetric action of the diffusion operator $e^{-t_1\epsilon \hat K}$.

Below we proceed with an explicit derivation of the combination of this scheme with the pair product ansatz (PPA) for the N-particle density matrix  introduced by Pollock and Ceperley~\cite{Ceperley1995} which was efficiently employed in the many RPIMC simulations~\cite{Brown_2014,Militzer_2000,Militzer2001,Hu2011,Militzer_2021}. 

Let us explicitly write the contribution of binary interactions in the N-body density operator of a two-component system 
composed of electrons (e) and ions (I) 
\begin{eqnarray}
e^{-\epsilon\hat H}=e^{-\epsilon (\hat K_e+\hat K_I+\hat V_{ee}+\hat V_{II}+\hat V_{eI})},
\end{eqnarray}
where each operator is the sum of one-particle or two-particle operators. 
Later we will explicitly use the large mass asymmetry of the species, $m_I/m_e\gg 1$.
Now we apply the fourth-order factorization result~(\ref{4thorder}) with redefined (non-commuting) operators
\begin{eqnarray}
&&\hat{\bar{K}}=\hat K_e+\hat K_I+\hat V_{eI},\\
&&\hat{\bar{V}}=\hat V_{ee}+\hat V_{\text{II}}\,,
\end{eqnarray}
and evaluate the additional quantum corrections, $\hat W_Q$, to the bare operator $\hat{\bar{V}}$ by Eq.~(\ref{W1})--(\ref{W}). After omitting all mutually commuting operators we are left with the final result 
\begin{eqnarray}
&&\hat{\bar{W}}_Q=\hat W_{ee,Q}+\hat W_{II,Q},\\
&&\hat W_{ee,Q}=\frac{\hbar^2}{m_e} \epsilon^2 \sum\limits_{i=1}^{N_e} \abs{\vec F_{e,i}^2},\\
&&\hat W_{II,Q}=\frac{\hbar^2}{m_I} \epsilon^2 \sum\limits_{j=1}^{N_I} \abs{\vec F_{I,j}^2},
\end{eqnarray}
where $\vec F_{e(I),i}$ is the full force acting on particle ``i'' only from the same particle species [i.e. with the exclusion of the e-i interaction].  
The corresponding matrix elements are diagonal in the coordinate representation
\begin{eqnarray}
&&\langle \vec R_{p-1} | e^{-\epsilon \hat{\bar{W}}_n} | \vec R_p \rangle=e^{-\epsilon W(\vec R_{p})}\,\delta(\vec R_p-\vec R_{p-1})\,,
\label{UD1}\\ 
&&W=-\frac{1}{2}\epsilon\,\left[ \bar{W}_n(\vec R_{p-1}; \tau_{p-1})+\bar{W}_n(\vec R_p; \tau_{p})\right],\\
&&\bar{W}_n=v_n(V_{ee}+V_{II})+u_0 a_n (W_{ee,Q}+W_{II,Q}).\label{UD2}
\end{eqnarray}
In particular, the $ee$-correlations are treated in the fourth-order factorization scheme the same as in the UEG case~\cite{dornheim_njp15,filinov_cpp_21}. As an alternative approach to Chin's result, Eqs.~(\ref{W1})--(\ref{W}), we can include first quantum corrections via the effective quantum potentials, e.g. the IKP (see Sec~\ref{ss:qpp}). Note that a direct use of the IKP in Eqs.~(\ref{W1})--(\ref{W}) is not justified as this would lead to double counting of quantum diffraction effects. 

Hence, for particles with the same charge and mass we  benefit from the fast $P$-convergence of  Chin's scheme in a one-component system~\cite{Sakkos2009,dornheim_njp15,filinov_cpp_21}. However, this scheme will fail for two electrons with different spins, once, at low temperature/high density, they closely approach each other, e.g. within a molecule or by scattering of two neutrals, cf. Fig.~\ref{fig:PDFT15P}c. This behavior will be prohibited by a divergent Coulomb force in Eq.~(\ref{W}), unless a tiny time-argument $\epsilon$ is employed. Therefore,
 for plasma simulations we use a hybrid scheme: $V_{ee}^{\uparrow\uparrow},V_{ee}^{\downarrow\downarrow}$ and $V_{II}$ are treated via Eqs.~(\ref{W1})--(\ref{W}), whereas $V_{ee}^{\uparrow\downarrow}$ is treated via the IKP~(\ref{Kelbg}) with the fit parameter $\gamma_{ee}(\epsilon,\mu_{ee})$ \cite{filinov_jpa03}.  

To complete our result, we proceed with the evaluation of the second operator (diffusion operators plus the electron-ion contribution). To shorten the notations we use $\epsilon \equiv t_n \epsilon\; (n=0,1)$:
\begin{eqnarray}
&&\rho(\vec R_{p-1},\vec R_{p},\epsilon)=\langle \vec R_{p-1} | e^{-\epsilon \hat{\bar{K}}} | \vec R_p \rangle \approx\nonumber \\
&& \int \db \vec R_p^\prime \,  \langle \vec R_{p-1} | e^{-\epsilon (\hat K_e+\hat V_{eI}}) | \vec R_p^\prime \rangle \langle \vec R_{p}^\prime | e^{-\epsilon \hat K_I} | \vec R_p\rangle.\label{rhoEI}
\end{eqnarray}
In the second line we employed a first approximation, and neglected the commutator term, $ \epsilon^2 [\hat V_{eI},\hat K_I]$, which is justified by the nearly classical behavior of ions. 

In the second step, we use the PPA, with the explicit result,  $|\vec R^{\prime\, e}_{p}\rangle=|\vec R_{p}^e\rangle$, which follows from the fact that  the diffusion operator (in the second term) acts only on the ion variables. This allows us to simply the first term to
\begin{eqnarray}
&&\langle \vec R_{p-1} | e^{-\epsilon (\hat K_e+\hat V_{eI})} | \vec R_{p}^\prime\rangle=\nonumber \\
&& \langle \vec R_{p-1}^I|\vec R_p^{\prime \, I}\rangle \times  \rho^e_0(\vec R_{p-1}^e,\vec R^e_{p},\epsilon)\, \times \nonumber \\
&&\prod \limits_{i=1}^{N_e} \prod \limits_{j=1}^{N_I} \exp{\left[- \epsilon\, U_{eI}(\vec r^{ij}_{p-1},\vec r^{ij}_p)\right]},\\
&&\vec r^{ij}_{p-1}=\abs{\vec r^{i,e}_{p-1}-\vec r^{j,I}_{p-1}}, \quad \vec r^{ij}_{p}=\abs{\vec r^{i,e}_{p}-\vec r^{j,I}_{p-1}}\,,
\end{eqnarray}
After performing the integral over $\vec R_p^\prime$, we get our final result for the DM in Eq.~(\ref{rhoEI})
\begin{eqnarray}
&&\rho(\vec R_{p-1},\vec R_{p},\epsilon)=\nonumber \\
&&\rho^e_0(\vec R_{p-1}^e,\vec R^e_{p},\epsilon)\times  \rho^I_0(\vec R_{p-1}^I,\vec R^I_{p},\epsilon) \nonumber\\
&&\times \prod \limits_{i=1}^{N_e} \prod \limits_{j=1}^{N_I} \exp{\left[- \epsilon\, U_{eI}(\vec r^{ij}_{p-1},\vec r^{ij}_p)\right]}\,,
\end{eqnarray}
where $\rho_0^e$ and $\rho_0^I$ are the $N$-body free-particle DM for electrons and ions, and $U_{eI}$ is the exact pair potential for the electron-ion problem. An explicit numerical scheme to evaluate $U_{eI}$ was proposed by Storer and Klemm~\cite{Storer1968,Klemm1973}, and a parametrization was introduced by Pollock~\cite{Pollock1988} and  Ceperley~\cite{Ceperley1995}. Its diagonal approximation via~(\ref{diagDM}), at high temperatures, converges to the IKP~\cite{filinov_pre04}, but is more accurate for $T\lesssim 1 Ry$.

The above derivation by its physical assumptions resembles the adiabatic approximation but applied at the ``elevated'' (via the factorization) higher temperature $T\sim \epsilon^{-1}$. The thermal fluctuations in the ion trajectory propagating from $| \vec R_p^I \rangle$ to $| \vec R_p^{\prime \, I} \rangle$ are induced by the diffusion operator $e^{- \epsilon \hat K_I}$, and are of the order of the ion DeBroglie wavelength~$\lambda_I(\epsilon)$, Eq.~(\ref{eq:lambda}). This length scale should be much smaller than the Bohr radius -- the characteristic spatial range of the e-i interaction.
In this case the interaction energy in the DM can be approximated in different ways, cf. $V_{eI}(\abs{\vec r^e_{p}-\vec r^{\prime\, I}_{p}})\approx V_{eI}(\abs{\vec r^e_{p}-\vec r^I_{p}})\approx V_{eI}(\abs{\vec r^e_{p}-\vec r^I_{p-1}})$.

In the final step, the free-particle electron DM should be antisymmetrized as in Eq.~(\ref{ZF2}). 
This can be done by splitting the e-i pair potential into a diagonal (D) and an off-diagonal (OD) contribution
\begin{eqnarray}
&&U_{eI}(\vec r_{p-1},\vec r_p; \epsilon)=\nonumber \\
&&\frac{1}{2}\left[u^D_{\epsilon}(\vec r_{p-1})+u^D_{\epsilon}(\vec r_p) \right]+ u^{\text{OD}}_{\epsilon}(\vec r_{p-1},\vec r_p).\label{UeIoffD}
\end{eqnarray}
By performing, as before, the summation over permutations in Eq.~(\ref{Zf}), we obtain the same representation for the partition function as in~(\ref{ZF2}) with the following modifications:

First, the number of imaginary time slices in the particle trajectories is changed $P \rightarrow 3P$, due to the 4th-order representation~(\ref{4thorder}) [we perform two additional diffusion steps weighted by the interactions $\hat W_{1(2)}$];

Second, the  anti-symmetric diffusion matrix 
now includes the off-diagonal part of the e-i pair potential directly in the matrix elements
\begin{eqnarray}
&&\mathbb M^{s,\text{int}}=||m_{kl}^{\text{int}}(p-1,p)||, \quad k,l=1,\ldots N^s\,,\label{M2} \\
&&m_{kl}^{\text{int}}(p-1,p)=\nonumber \\
&&\exp\left(-\frac{\pi}{\lambda^2_{\epsilon}}[\vec r^s_{l,p}-\vec r^s_{k,p-1}]^2-\epsilon \sum\limits_{j=1}^{N_I} u^{\text{OD}}_{\epsilon}(\vec r_{p-1}^{k j},\vec r_{p}^{l j})\right)\,,  \nonumber 
\end{eqnarray} 
where the sum runs over all ions, and the vectors $\{\vec r_{p-1}^{k j}$,$\vec r_{p}^{l j}\}$ specify the e-i distances on two successive time slices. Note, that in the Slater determinant all electron coordinates are permuted, and each electron does not need to be specified by a distinguished trajectory, as in PIMC for Bose systems~\cite{Ceperley1995} or in RPIMC~\cite{Ceperley1991}; 

Third, the potential energy part is now diagonal and includes several contributions [the arguments are defined as $x=(\vec R_{p-1},\vec R_{p-1}; t_n \epsilon)$ and $x^\prime=(\vec R_{p},\vec R_{p}; t_n \epsilon)$]:
\begin{enumerate}

    \item Due to 4th-order factorization,  
\begin{eqnarray}
&&U^{\text{4th}}(x,x^\prime)=\frac{1}{2}\left[W(x)+W(x^\prime)\right],\label{U1}\\
&&W(x)=W_{ee}^{\uparrow\uparrow}(x)+W_{ee}^{\downarrow\downarrow}(x)+W_{\text{II}}(x),
\end{eqnarray}
\item The IKP potential for the opposite spin electrons
\begin{eqnarray}
U_{ee}^{\uparrow\downarrow}(x,x^\prime)=\frac{1}{2}\left[U_{ee}^{IKP}(x)+U_{ee}^{IKP}(x^\prime) \right],\label{UeeIKP}
\end{eqnarray}
or, alternatively, we use the diagonal part $u_\epsilon^D$ of the exact pair potential for two electrons, similar to Eq.~(\ref{UeIoffD}). 
\item The e-i interaction is included via
the diagonal part of the e-i pair potential
\begin{eqnarray}
 U^{\text{D}}_{eI}(x,x^\prime)=\frac{1}{2} \left[u^D(x)+u^D(x^\prime) \right].\label{U2}
\end{eqnarray}
\item The off-diagonal contributions to the ei-interaction are explicitly included in the Slater determinants, and account for the indistinguishable nature of the same spin electrons.  
\end{enumerate}

Note, that only the 4th-order factorization includes the bare Coulomb potential, whereas the other effective  potentials, $U^{\uparrow\downarrow}$ and $U_{eI}$, are soft and permit formation of bound states even with only a few high-temperature factors involved, e.g. $P\sim 3$.

%$|\vec R_{p}^{\prime \, I}\rangle =|\vec R^{I}_{p-1}\rangle$, and correspondingly, $|\vec R^\prime_{p}\rangle =| \vec R_p^e \vec R^{I}_{p-1} \rangle$.
%The operator of kinetic energy for the ions take the form

With the above derivation, we have achieved our main goal:  exchange effects related to the electron spin are taken into account via the anti-symmetric short-time propagators~(\ref{M2}) in the form of Slater determinants modified by e-i correlations.

\subsection{Thermodynamic functions}\label{Estim}
In this section we present the estimators for the main thermodynamic functions of interest used in our FP-PIMC approach.
We start from the definition of the partition function,
\begin{eqnarray}
 Z=\int \db \vec{X} \int \prod\limits_{p=1}^P e^{-\epsilon_p U(\vec R_p)} \det \mathbb M[\rho_0(\vec R_{p-1},\vec R_p;\tau)]\,, \nonumber
\end{eqnarray}
where, $\vec R_p=(\vec r_{1,p},\vec r_{2,p},\ldots,\vec r_{N,p})$, contains the coordinates of all particles. Compared to the exact representation, cf. Eqs.~(\ref{ZF2}),~(\ref{UD2})--(\ref{M2}), and (\ref{W1})--(\ref{W}) we have simplified the notations to highlight the general structure of the related thermodynamic estimators. For example, the single determinant represents
 the contributions of two Slater determinants (for the spin up/down electrons). The explicit expressions of the total, kinetic and potential energy for the 4th-order factorization is re-derived with additional contributions  from the quantum correction, $\hat W_Q$~(\ref{W}), for details see Sakkos {\it et al.}~\cite{Sakkos2009}. Also the summation over $P$ should be extended to $3P$ due to additional factorization factors in~(\ref{4thorder}).

%In particular, the virial estimator of the pressure will require the estimation of the tensor from the force-term, $\hat W_Q$.  

The total interaction energy $U$ is  the sum of the pair potentials of particles of the same kind and between different species, 
\begin{eqnarray}
 U(\vec R_p)=\sum_{\al} \sum_{i<j}^{N_\al} u_{\al\al}(\vec r^{ij}_p)+\sum_{\al} \sum_{\be} \sum\limits_{i=1}^{N_\al}\sum\limits_{j=1}^{N_\be}  u_{\al\be}(\vec r^{ij}_p)\,.
\end{eqnarray}
\subsubsection{Pressure}\label{ss:pressure}
The pressure is related to the partition function via 
\begin{eqnarray}
 \beta p = \frac{\partial \ln Z}{\partial V}\,,
\end{eqnarray}
where the derivative is performed by  introducing a scaling factor, $\alpha=L/L_0=(V/V_0)^{1/3}$, and by re-scaling all particle coordinates, $\vec r \rightarrow \alpha \vec r$,
\begin{eqnarray}
 \frac{\partial \ln Z}{\partial V}=\left[\frac{\partial \ln Z(\al \vec{X})}{\partial \alpha} \frac{\alpha}{3 V} \right]_{\alpha=1}.
\end{eqnarray}
This standard procedure, applied within the FP-PIMC representation of the $N$-body DM, leads to several contributions which originate from its kinetic and potential parts, as well as  additional terms, in the case of quantum potentials, such as the (improved) Kelbg potential,
% where both contributions are mixed in the effective potential function, see 
Eq.~(\ref{Kelbg}). 

The interaction-induced contribution to the pressure is obtained as the $\al$-derivative of the potential energy term with the substitution $\{\vec x_{i} =\alpha \vec r_{i},\;\; i=1,\dots N\}$
\begin{eqnarray}
 \frac{\partial U(\vec X_p)}{\partial \alpha}=\sum\limits_{i<j} \sum\limits_{d=1}^3   \left[\frac{u(\vec x_p^{ij})}{\partial x_p^{ij,d}} \frac{\partial x_p^{ij,d}}{\partial \alpha}\right]_{\alpha=1}=\sum\limits_{i<j}^N \nabla u(\vec r_p^{ij}) \cdot \vec r_p^{ij} \nonumber  
\end{eqnarray}
and leads to an analogous result as in classical Statistical Physics, with an additional averaging along the particle trajectories
\begin{eqnarray}
  p^{V}=-\frac{1}{3V} \left\langle \frac{1}{P}\sum\limits_{p=1}^P \sum\limits_{i<j}^N \nabla u(\vec r_p^{ij}) \cdot \vec r_p^{ij}\right\rangle\,,
\end{eqnarray}
with $\vec r^{ij}_p=\vec r_{i,p}-\vec r_{j,p}$, where the coordinate vectors $\vec r_{i,p}=\vec r_i(\tau_p)$ depend on the imaginary time argument, $0\leq \tau_p \leq \beta$, with $1\leq p\leq P$.

The long-range interactions with the periodically repeated main simulation cell are evaluated either via the Ewald summation or the angle-averaged Yakub potential (``Y'') \cite{yakub_ltp_05} which gives rise to additional contributions.
%\textcolor{red}{
We have tested both potentials and found that, for the simulation parameters used in this paper, the results for both are practically indistinguishable.
%}
%{\color{blue} Both have been used. I will add a figure which compares both potentials}
For the case of the Yakub potential the long-range interaction term
\begin{eqnarray}
u_t^Y(\vec r^{ij}_p)=\frac{2 \pi}{3 V} r_p^2,
\end{eqnarray}
 gives rise to the {\it tail correction} to the pressure 
\begin{eqnarray}
p_1^{Y}= \left\langle \frac{1}{P}\sum\limits_{p=1}^P \sum\limits_{i<j}^N \frac{\partial  u^{Y}_t(x^{ij}_p)}{\partial \alpha}|_{\al=1}\right\rangle=\left\langle-\left[\frac{3}{\alpha}u^{Y}_t(r^{ij}_p)\right]_{\alpha=1} \right\rangle \nonumber
\end{eqnarray}
and an additional contribution to the virial part
\begin{eqnarray}
  p_2^{Y}=\frac{1}{V} \left\langle \frac{1}{P}\sum\limits_{p=1}^P \sum\limits_{i<j}^N \nabla u^Y_t(\vec r^{ij}_p) \cdot \vec r^{ij}_p)\right\rangle\,.
\end{eqnarray}
Next, we take into account the contribution of the kinetic energy part (constructed with the Slater determinants between adjacent time slices $\{\tau_{p-1},\tau_{p}\}$) 
\begin{eqnarray}
 &&p^{T}=\frac{1}{3V} 3 N k_BT\cdot P+\\ \nonumber
 &&\left\langle\frac{1}{3V P}\sum\limits_{p=1}^P \frac{\partial \det \mathbb M[\rho(\al\vec R_{p-1},\al\vec R_p] /\partial \alpha}{\det \mathbb M[\rho(\vec R_{p-1},\vec R_p]}\right\rangle\nonumber\,.
\end{eqnarray}
Alternatively, using the result for the $\al$-derivative of the matrix elements in $\mathbb M[m_{kl}]$
\begin{eqnarray}
&&\left[ \frac{\partial m_{kl}(\al\vec r_{k,p-1},\al\vec r_{l,p})}{\partial \alpha}\right]_{\al=1}=m_{kl,p}^{0,\al},\\
&&m_{kl,p}^{0,\al}=-\frac{2}{\lambda^2_{\tau_p}}\left(\vec r_{l,p}-\vec r_{k,p-1}\right)^2 \, e^{-\frac{(\vec r_{l,p}-\vec r_{k,p-1})^2}{\lambda^2_{\tau_p}}} \nonumber
\end{eqnarray}
the same contribution is directly related to the thermal part of the total kinetic energy, $E_{\text{kin}}^{T}$, as follows \begin{equation}
p^{T} =\frac{2}{3V} E_{\text{k}}^{T}.\label{PT}
\end{equation}
Once, the e-i interactions are included in the matrix elements via Eq.~(\ref{M2}), we get an additional contribution related to the off-diagonal potential
\begin{eqnarray}
&&\left[ \frac{\partial m^{\text{int}}_{kl}(p-1,p)}{\partial \alpha}\right]_{\al=1}=m^{\text{int},\al}_{kl,p}\\
&&m^{\text{int},\al}_{kl,p}/m^{\text{int}}_{kl}(p-1,p)=-\frac{2}{\lambda^2_{\tau_p}}\left(\vec r_{l,p}-\vec r_{k,p-1}\right)^2  \nonumber \\
&&-\epsilon \sum\limits_{j=1}^{N_I} \frac{\partial}{\partial \al} u^{\text{OD}}_{\epsilon}(\al\vec r_{p-1}^{k j},\al\vec r_{p}^{l j})|_{\al=1} \,.
\end{eqnarray}

Thus, the full  pressure estimator 
is given by 
\begin{eqnarray}
p=p^{T}+p^{V}+p_1^{Y}+p_2^{Y}\,. \label{virialP}
\end{eqnarray}

%Derivative of the matrix elements
%\begin{eqnarray}
% \frac{\partial \rho_{ij}(\vec r^k_i,\vec r^{k+1}_j)}{\partial \alpha}=-\frac{2}{\lambda^2_{\tau_k}}(\vec r^{k+1}_j-\vec r^{k}_i)^2 \, e^{-(\vec r^{k+1}_j-\vec r^{k}_i)^2/\lambda^2_{\tau_k}}
%\end{eqnarray}

\subsubsection{Kinetic energy}\label{ss:kin-energy}
The PIMC representation of the full kinetic energy is based on the exact  estimator
\begin{eqnarray}
E_k=\sum_{\al}\sum_{i=1}^{N_{\al}}\frac{m_i^{\al}}{\beta}\; \frac{\partial Z}{\partial m_i^{\al}}\,.
\label{Ekin}
\end{eqnarray}
In our case the mass derivative of the matrix elements in the free-particle Slater determinants can be sequentially reduced, first, to the partial derivatives with respect to the two-particle reduced de Broglie wavelength, $\lambda_{ij}$, see Eq.~(\ref{Kelbg}), and, in a second step, to the $\al$-derivatives used in Sec.~\ref{ss:pressure} for the pressure estimator. This way we can directly prove Eq.~(\ref{PT}) and write the intermediate result
\begin{eqnarray}
&&E_k^T=\frac{3}{2} N k_B T P+ \nonumber \\ 
&&\left\langle\frac{1}{2P}\sum\limits_{p=1}^P \frac{\partial \det \mathbb M[\rho(\al\vec R_{p-1},\al\vec R_p] /\partial \alpha}{\det \mathbb M[\rho(\vec R_{p-1},\vec R_p]}\right\rangle\,.
%+E_{\text{kin}}^{\lambda}
\end{eqnarray}
Some additional care should be taken with the use of effective quantum potentials. 
Both, the exact pair and the improved Kelbg potentials, cf. $U_{ee}^{IKP}$ and $U_{eI}^D$, Eqs.~(\ref{UeeIKP}) and (\ref{U2}), contain an additional dependence on the temperature and particle masses via $\lambda_{ij}$. This leads to an additional contribution to the kinetic energy and also to the total energy.
Using Eq.~(\ref{Ekin}) we can estimate the corresponding correction 
\begin{eqnarray}
&&E_{k}^{\lambda}=\left\langle\frac{1}{2P}\sum\limits_{p=1}^P \sum\limits_{i,j}^N  \lambda_{ij} \frac{\partial u(r^{ij}_p)}{\partial \lambda_{ij}}\right\rangle\,,\\
&& u(x)=U_{ee}^{\uparrow\downarrow}(x)+U^D_{eI}(x)\,.
\end{eqnarray}
%where the correct selection of binary interactions in $\sum_{i,j}^N$ (for each case -- ee, eI ) is assumed. 
 %
Finally, we express the full kinetic energy as
\begin{eqnarray}
 E_{k}=E_k^T+E_{k}^{\lambda}.\label{Ek_final}
\end{eqnarray}

\subsubsection{Total and potential energy}
The total energy is given by 
%the $\be$-derivative of the partition function
\begin{eqnarray}
E_T=-\frac{1}{Z}\frac{\partial Z}{\partial \beta}\,.
\end{eqnarray}
As the inverse temperature $\beta=\tau P$ is directly contained in $\lambda_{ij}(\tau)$, the kinetic energy case, discussed above, can be directly used, with the result
\begin{eqnarray}
E=E_k^T+E_{k}^{\lambda}+ E_p
%\left\langle\frac{1}{P}\sum\limits_{p=1}^P \sum\limits_{i<j}^N  u(r^{ij}_p)+u^Y_t(r^{ij}_p)\right\rangle
\;.
\end{eqnarray}
The potential energy follows directly from Eq.~(\ref{Ek_final})
\begin{eqnarray}
E_p=E-E_k=\left\langle\frac{1}{P}\sum\limits_{p=1}^P \sum\limits_{i<j}^N  u(r^{ij}_p)+u^Y_t(r^{ij}_p)\right\rangle  + \Delta E^{\rm OD}_p\,.\; \label{ep}
\end{eqnarray}
where $\Delta E^{\rm OD}_p$ denotes the correction due to the off-diagonal quantum pair potential $u^{\rm OD}_\epsilon$.

\section{Simulation results}\label{s:results}

We have carried out fermionic PIMC calculations in a broad range of densities and temperatures relevant for warm dense matter. The simulations have been performed for a deuterium plasma (D), with $m_D/m_e=3673$, but our data are applicable to the hydrogen (H) EOS as well by a simple rescaling of the mass density via the relations,
\begin{eqnarray}
&&\rho_{\text{D}}[\text{g/cm}^{3}]=(1.75313/r_s)^3, \\
&&\rho_{\text{H}}[\text{g/cm}^{3}]=(1.39181/r_s)^3,   
\end{eqnarray}
and $\rho_{\text{D}}=1.9985\; \rho_{\text{H}}$. 
%The same equations can be used to recalculate the mass density from the $r_s$-values. 
In the discussion of the results in the following sections we will mainly refer to hydrogen density $\rho_H$.

The EOS was evaluated 
for a series of isotherms covering the range $15,600$~K $\leq T \leq 400,000$~K.
The analyzed density interval ($1 \leq r_s \leq 100$) for hydrogen (deuterium) corresponds to $7 \cdot 10^{-7}$g/cm$^{3}\leq \rho_H \leq 0.085$ g/cm$^{3}$ ($1.4 \cdot 10^{-6}$g/cm$^{3}\leq \rho_D \leq 0.17$ g/cm$^{3}$) and  
 to electron degeneracy parameter values  of $\theta \gtrsim 0.5$ and  $\chi  \lesssim 3$. 

For all simulation conditions the collected data for the pressure and the internal energy isotherm are presented in the combined EOS-table, see Tab.~\ref{tab:data} in Appendix~\ref{eos-table}. The statistical errors strongly depend on the temperature-density combination and the corresponding values of the degeneracy parameters $\{\theta,\chi\}$,  which influences the severity of the FSP in our simulations. 
%The attenuation of the average sign~(\ref{def_Sign}) in some cases can significantly enhance the statistical error.

In addition, we resolve the kinetic and the potential energy contributions, and this way we are able to verify the virial theorem in our thermodynamic data. In particular, we employed the {\it thermodynamic}~(\ref{virialP}) and the {\it virial} (valid for Coulomb systems)
\begin{eqnarray}
    p^v V= \frac{2}{3} E_k+ \frac{1}{3} E_p,\label{Pvir}
\end{eqnarray}
estimators for the pressure. Both are found to agree within the statistical errors if the number $P$ is sufficiently high.
%
%Based on the thermodynamic functions presented in Sec.~\ref{Estim}, a first-principle EOS (Tab.~\ref{Tab-EOS}) is reconstructed in a wide range of parameters and compared to other theoretical many-body techniques.

\subsection{Convergence analysis}\label{ss:convergence}
To validate the numerical accuracy of our simulations
we performed a convergence analysis of the main thermodynamic quantities with respect to the number $P$ of high-temperature factors in the fermionic partition function $Z_F$, Eq.~(\ref{ZF2}), which was done for    
 three representative temperatures, $T=15,640$~K, $31,250$~K, and $125,000$~K. 

\subsubsection{Convergence of the pair distribution functions (PDF)}\label{sss:pdf}
\begin{figure}[]
\hspace{-.8cm}
\includegraphics[width=0.53\textwidth]{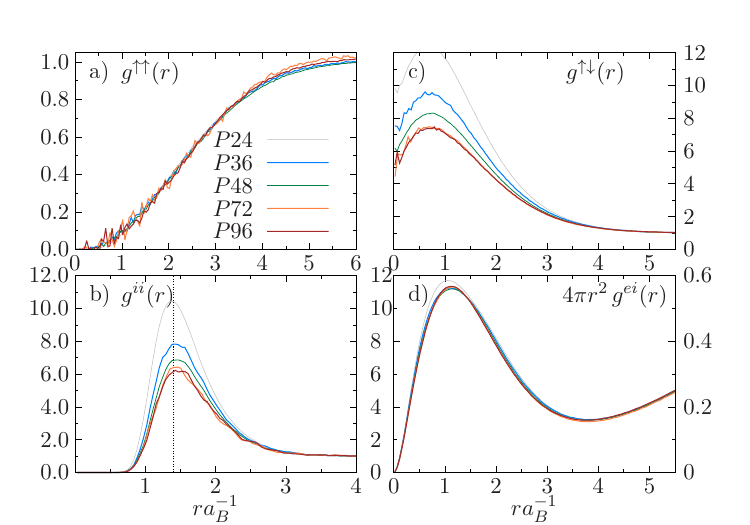}
\caption{Dependence of the pair distribution functions on the number of factors $P$, for $T=15,640$~K, $r_s=7$, and the particle number $N=34$. a) PDF for same spin electrons; b) ion PDF; c) PDF for electrons with anti-parallel spins; d) 
electron-ion PDF. The presence of neutral bound complexes $H, H_2$ and molecular ions $H_2^+$ can be identified from the broad peaks in $g^{ii}(r)$ for ion-ion distances, $1.1 a_B \leq r \leq 2.2 a_B$, and $g^{\uparrow\downarrow}(r)$. The fraction of molecules, contributing to the peak height in $g^{ii}(r)$ (the vertical dotted line denotes the bond length $l_B=1.4 a_B$ in $H_2$), however, is strongly overestimated for the cases with only few propagators $P=24(36)$. Full convergence is achieved only at $P=72(96)$. The atom fraction (related to the peak height in panel d) has only a weak $P$-dependence.}
\label{fig:PDFT15P}
\end{figure}

\begin{figure}[h]
\hspace{-.8cm}
\includegraphics[width=0.52\textwidth]{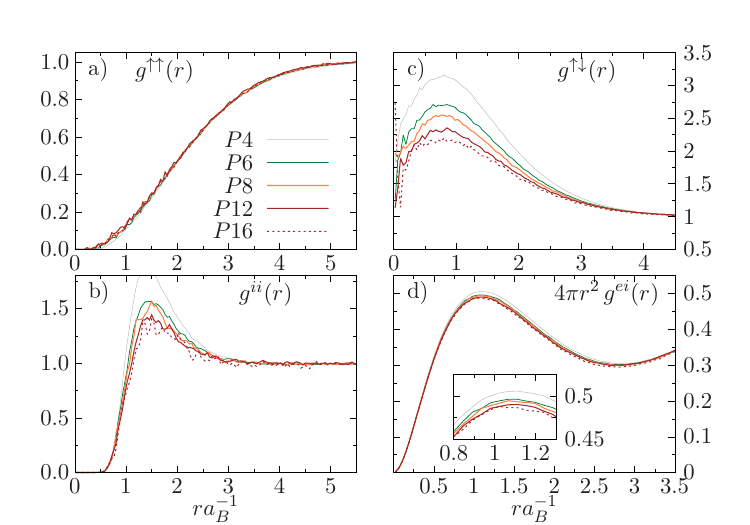}
\vspace{-0.5cm} 
\caption{Same as Fig.~\ref{fig:PDFT15P}, but for $T=31,250$~K and $r_s=5$.  The molecular fraction is strongly overestimated for $P=4$, and  convergence is achieved at $P=12(16)$, [cf. Fig. b)]. As for lower temperatures, the peak height in $ 4 \pi r^2 g^{ei}$ in panel d) exhibits only a weak $P$-dependence.}
\label{fig:PDFT31P}
\end{figure}
First, we address the $P$-convergence of the PDF 
at $T=15,640$K, in Fig.~\ref{fig:PDFT15P}, and $T=31,250$K, in Fig.~\ref{fig:PDFT31P}.  
The four panels clearly demonstrate how sensitive the inter-particle correlations  are to the choice of $P$. In PIMC simulations higher $P$-values are, in general, required to accurately include the effects of three-body and higher-order correlations. This is obviously the case for the attractive Coulomb interaction experienced by the electrons within an atom or a molecule what explains the significantly slower convergence of both $g^{\upspin\downspin}(r)$ and $g^{ii}(r)$, compared to $g^{\upspin\upspin}(r)$ and $g^{ei}(r)$. 

The increase of $g^{\upspin \downspin}>1$, below $r=1a_B$ (panels b) indicates the formation of  molecules, which are also clearly observed 
as attractive correlations between  pairs of atoms,  with a broad peak in $g^{ii}(r)$ at $r\sim 1.4 a_B$ (panels c).
Note, that the peak amplitude becomes strongly suppressed at $P > 48(15kK)$, and $P> 12$ (31kK). Convergence is reached only for $P\ge 72$, for $15$kK and $P\ge 16$, for $31$kK (the corresponding  lines agree within the statistical errors). The same trend is observed in $g^{\upspin\downspin}(r)$ as well. By using too low $P$-values [e.g. $P=24$, for 15kK, and $P=8$, for 31kK] the attractive interaction within the bound complexes ($H_2, H_2^+$) is significantly overestimated. First, this has a strong effect on the plasma composition at low temperatures (see the cluster analysis in Sec.~\ref{lowT}), and, second, influences all thermodynamic functions including the EOS.

In contrast, significantly lower $P$-values are sufficient to describe the correlations between the spin-like electrons (panels a). These electrons do not participate in molecule formation and, their short-range Coulomb repulsion is enhanced by 
Pauli blocking which is taken into account by the anti-symmetrization of the many-body density matrix employing Slater determinants~(\ref{M2}). A similar effect was observed for the UEG~\cite{Groth_2016,filinov_cpp_21}, where only few (fourth-order) propagators [$P\sim 2\dots 3$] were found to be sufficient even for temperatures  below the Fermi temperature, $0.5 \le \theta$. 

Finally, we analyze the electron-ion PDF (panels d). It does show some weak $P$-dependence, but only in the range $0.5 a_B\leq r \leq 1.5 a_B$. This behavior is physically reasonable: at short distances ($r\leq 0.5 a_B$) pairwise e-i correlations dominate over the many-body effects, and the employed pair approximation for the $N$-body density matrix~\cite{RevModPhys.67.279,Militzer_2000} becomes nearly exact. 
The influence of the plasma environment on the electron density within an atom becomes significant only at distances $r\gtrsim a_B$ where the pair-product ansatz for the $N$-body density matrix is not appropriate. 
\begin{figure}[]
\begin{center}
\hspace{-0.4cm}
\includegraphics[width=0.52\textwidth]{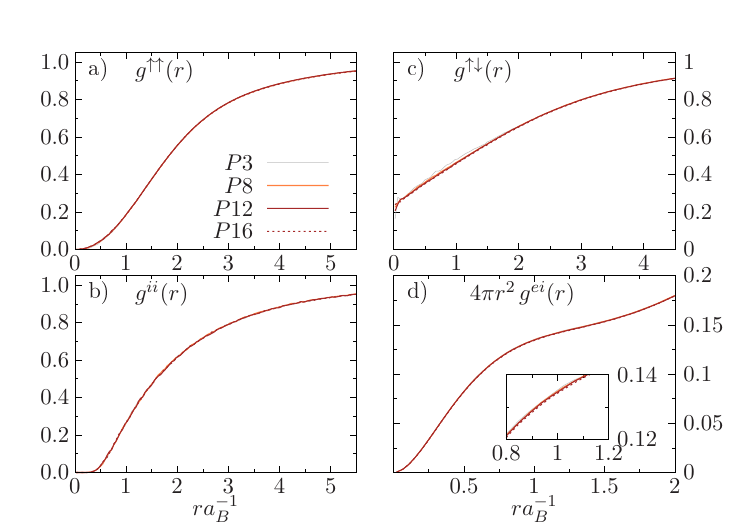}
\vspace{-0.3cm} 
\caption{The PDF similar to Fig.~\ref{fig:PDFT15P} at $T=125,000$~K and $r_s=5$. In contrast to low temperatures ($T \leq 35, 000$K),  $P=3$ is completely sufficient to accurately capture the electron-electron and the electron-ion correlations.}
\label{fig:PDFT125P}
\end{center}
\end{figure}

The relative importance of many-body effects for the PDF and the influence of the number $P$ is further analyzed in Fig.~\ref{fig:PDFT125P} for a higher temperature ($T=125,000$~K) when only few bound complexes are present. 
%The  $P$-convergence is illustrated in four panels in Fig.~\ref{fig:PDFT125P}. 
For the density 
%$\rho \sim 0.02$ g cm$^{-3}$ 
$r_s=5$ the lowest value $P=9$ is completely sufficient for accurate thermodynamic functions. Some minor effects can be resolved only in the e-i PDF (panel d) at $r\sim a_B$, as demonstrated in the inset. Physically, a particular choice of $P$ has an effect on the interaction between neutrals and free electrons once they approach each other to distances comparable to the effective atomic radius. This effect can be observed as an onset of formation of a local maximum in $4 \pi r^2 g^{ei}$ at $0.8\lesssim r a_B^{-1}\lesssim 1.2$, and this type of correlations exhibits the slowest $P$-convergence. 

\subsubsection{Convergence of thermodynamic functions}\label{TFconverg}
\begin{figure}
\includegraphics[width=0.55\textwidth]{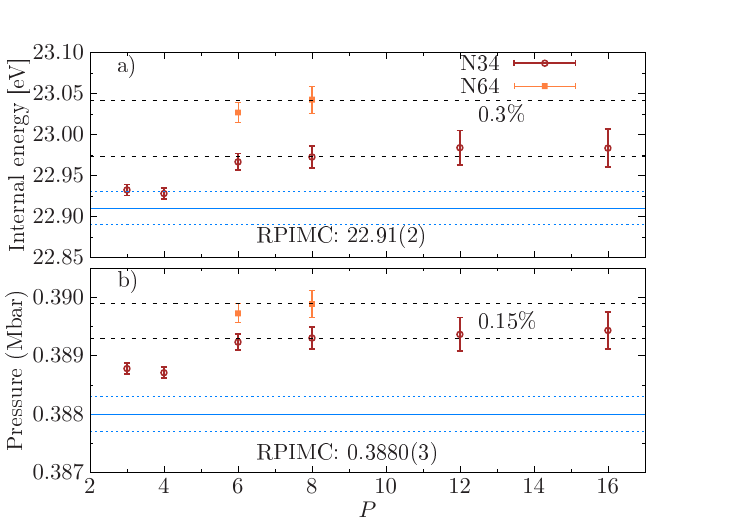}
\vspace{-0.5cm} 
\caption{
Convergence of (a) internal energy (per atom) and (b) pressure vs. $P$, for $r_s=5$ and $T=125, 000$K.
%and $N_{i(e)}=34,64$. 
The extrapolation to the $P\to \infty$ limit is indicated by the dashed lines. The numbers in percent indicate the relative deviation between two system sizes $N=34$ and $N=64$. The RPIMC data~\cite{Hu2011} are shown by the solid blue line (dashed lines are the error bars).}
\label{fig:covergT125}
\end{figure}
\begin{figure}[]
\includegraphics[width=0.5\textwidth]{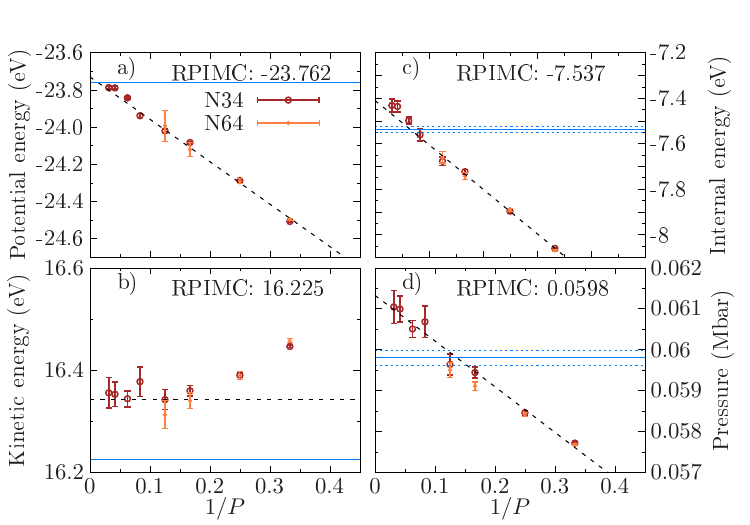}
\vspace{-0.5cm} 
\caption{$1/P$-dependence of (a) kinetic energy,  (b) internal energy, (c) kinetic energy , (d) pressure for $T=31, 250$~K and $r_s=5$ and a number of factors $3\leq P\leq 32$. 
The extrapolation to the $P\to \infty$ limit is indicated by the dashed lines. Horizontal blue lines: RPIMC data~\cite{Hu2011}.}
\label{fig:covergT31}
\end{figure}
For an accurate reconstruction of the EOS the $P$-convergence of the main  thermodynamic functions has to be verified, in addition to the behavior of the PDF.  
We start with the simplest case of high temperature, cf. Fig.~\ref{fig:covergT125}, when convergence of the PDF is achieved with a few factors,  $P=3$, see Fig.~\ref{fig:PDFT125P}.  
Similar conclusions can be drawn here. For $P\geq 6$, the FP-PIMC estimators for the energy and the pressure converge within the statistical error bars to the $P \rightarrow \infty$-limit. To demonstrate the finite-size effects we compare simulations for $N=34$ and $N=64$. The relative deviations in percent are cited in each panel. 
 
A similar analysis is presented  in Fig.~\ref{fig:covergT31}, for $T=31, 250$K and $r_s=5$. Both, the internal energy and the pressure converge to the asymptotic value, now for much larger $P$-values ($P\geq 24$).
A significant increase of the statistical error bars at high $P$ is due to the decay of the average sign~(\ref{def_Sign}) to $\Avr{S}\sim 0.035$  $(0.025)$, for $P=12$ $(P=16)$. Note, that the thermodynamic estimator
for the potential energy $\epsilon_p$ (panel a) exhibits a much slower $P$-convergence compared to the kinetic energy $\epsilon_k$ (panel b). This can be explained by a relatively slow $1/P$-convergence of atomic and molecular fractions (see Sec.~\ref{sss:composition-vs-p}), and a change in plasma composition has a stronger effect on $\epsilon_p$. The estimated molecular fraction, see Fig.~\ref{fig:Nfrac}b, is not negligible and reaches $5\%$ for $r_s=5$. The results saturate only for $P=24(32)$ propagators. The simulations for a larger system size $N=64$ ($P\leq 8$) agree within the error bars. 

After having established the convergence of the thermodynamic functions with $P$, we now compare the results to the RPIMC data [for details, see Sec.~\ref{ss:hydrogen-results}] which are shown in Figs.~\ref{fig:covergT125} and~\ref{fig:covergT31} by the blue lines with error bars. First, for $T=125, 000$K,  we observe  good agreement between FP-PIMC and RPIMC, where the deviations are about $0.6\%$, for the total energy and $0.5\%$, for the pressure ($N=64$), with the FP-PIMC data are being larger.  The same behavior of internal energy and pressure is observed for $T=31, 250$K, see the results of the $P\rightarrow \infty$ limit in Fig.~\ref{fig:covergT31}. From panel (b), it becomes clear that the main source of discrepancy is 
the kinetic energy which is underestimated in RPIMC by $0.1$eV which directly influences the pressure, due to the virial relation~(\ref{Pvir}). 
%Since the present FP-PIMC approach does not involve the fixed node approximation as used in RPIMC EOS~\cite{Hu2011}, and we have achieved convergence with respect to $P$,
%the revealed discrepancies call for a revision of the later approach.
%deviations are due to the inaccuracies of the latter.
%we can confidently validate the applicability of the fixed-node approximation~\cite{ceperley_alder,Ceperley1991,CeperleyFermi}, at least, in the high-temperature plasma phase. 
Some deviations to the RPIMC EOS become more noticeable at lower temperatures, and a systematic comparison, in a wide range of densities and temperatures, will be discussed in the following sections.

\begin{figure}[]
\hspace{-0.9cm}
\includegraphics[width=0.51\textwidth]{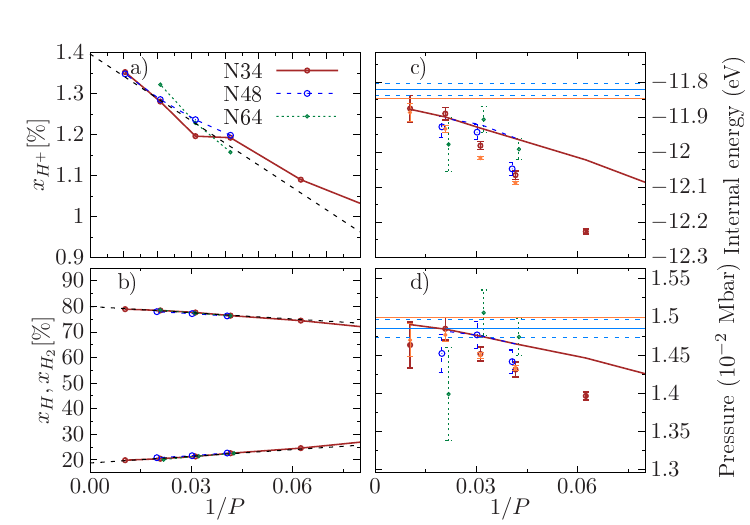}
\vspace{-0.2cm} 
\caption{P-convergence of the FP-PIMC results, for $r_s=6$, $T=15, 640$~K and  $N=34$ ($P\leq 96$), $N=64$ ($P\leq 72$, see different symbols). (a) Fraction of free ions. (b) Upper (lower) curve: fraction of atoms (molecules). (c) Internal energy, and (d) pressure. Solid brown ($N=34$), dashed blue ($N=48$) and dotted green ($N=64$) lines in (c) and (d) are the HSCM model results, Eqs.~(\ref{eq:e-hscm}, \ref{eq:p-hscm}),   with the $N$- and $P$-dependent fractions $\{x_{H^+},x_{H},x_{H_2}\}$ derived from the FP-PIMC data in panels (a) and (b). 
The extrapolation to the $P\to \infty$ limit is indicated by the horizontal (sienna) line. Horizontal blue lines in c) and d): RPIMC data~\cite{Hu2011}.
}
\label{fig:convT15rs6}
\end{figure}

\begin{figure}[]
\hspace{-0.9cm}
\includegraphics[width=0.51\textwidth]{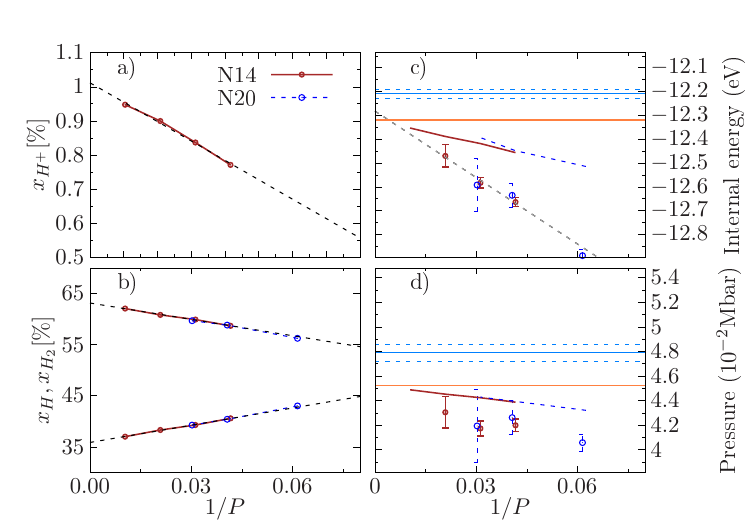}
\vspace{-0.4cm} 
\caption{Same as in Fig.~\ref{fig:convT15rs6}, but for $r_s=4$, $T=15, 640$~K and  $N=14,20$. Horizontal blue lines in c) and d): RPIMC data~\cite{Hu2011}.
}
\label{fig:convT15rs4}
\end{figure}

%significantly deviate from the RPIMC EOS data of Ref.~\cite{Hu2011}, $E=-7.537$eV, $P=0.0598$Mbar; 
%the relative difference 
%constitutes $14\%$, in the internal energy, and $8\%$ in the pressure, respectively. 
%These discrepancies cannot be explained by the discretization error induced by the finite $P$-value of our simulations. Another possible source of the deviations could be finite size effects. However, this can be ruled out, as will be discussed in the next subsection.

\subsubsection{Convergence of the plasma composition}\label{sss:composition-vs-p}
Next, in Figs.~\ref{fig:convT15rs4}--\ref{fig:convT15rs14} we concentrate on the low temperature case ($T=15,640$K), where the plasma is dominated by atoms and molecules. This regime was found to be the most difficult for the convergence analysis. A full $P$-convergence of all quantities cannot be achieved directly,  even with a number of $P=96$ fourth-order propagators, corresponding to approximately $300$ imaginary time slices in total. The main reason is the relatively slow convergence of the ion-ion and the opposite spin electron PDFs, see Fig.~\ref{fig:PDFT15P}. The peak height of both characterizes the change in the molecular (atomic) fraction and exhibits a strong $P$-dependence. As a result the bound electrons and ions contribute very differently to the kinetic and potential energy, depending on whether they belong to a molecule or to an atom. Also the dissociation equilibrium between the neutral bound complexes has a significant effect on the pressure. 

Taking into account these preliminary considerations, we develop a new scheme to perform an extrapolation to the $P\rightarrow \infty$ limit. It is based on a hard-sphere chemical model (HSCM), introduced in Appendix~\ref{ss:chem-model}, based on the numerical solution of the coupled Saha equations for the hydrogen ionization--dissociation equilibrium.
The corresponding thermodynamic expressions for pressure and energy of the model are defined by the equations
\begin{eqnarray}
   &&\beta p=n_a+n_m+n_i+\beta (p_e^{id}+p^{(3)}_{ex}+p_{ex}^C),
   \label{eq:p-hscm}\\
   && E=\sum\limits_{s=a,m,i,e}E_{id}^s+E_{ex}^{HS}+E_{ex}^C\,,
   \label{eq:e-hscm}
\end{eqnarray}
where $n_a$ and $n_m$ are the densities of atoms and molecules, $p_e^{id}$ is the pressure of an ideal Fermi gas, $E^s_{id}$ are the ideal energy contributions, and the definitions of the excess (ex) contributions are provided in Appendix~\ref{ss:chem-model}.

As the input the above equations require the species fractions $\{x_{H^+},x_{H},x_{H_2}\}$ resolved either from the HSCM or 
extracted from the FP-PIMC cluster analysis, $x_{i}=f_i(r_s,T,P,N)$, and contain an explicit dependence on the simulation parameters. 
In our simulations we have accurately determined the $P$-dependence of $\{x_{H^+},x_{H},x_{H_2}\}$ of free ions, atoms and molecules, as a function of $r_s$, $T$, and the system size $N$. These fractions have been analyzed at three different densities corresponding to $r_s=6,4$ and $14$ and are presented in Figs.~\ref{fig:convT15rs6}--\ref{fig:convT15rs14} (panels a, b). The deduced HSCM results for internal energy and pressure, Eqs.~(\ref{eq:e-hscm}) and (\ref{eq:p-hscm}), are compared in panels c), d) with the FP-PIMC data (symbols with  error bars). In each case, we observe that all fractions $x_s$  follow a $1/P$-scaling law, where $x_{H^+}$ [$x_H, x_{H_2}$] changes approximately by $0.5\%$ ($1\%$), from $P=96$ to $P\rightarrow \infty$, cf. panels a) [b)].  However, before using the extrapolated values, we have to verify that finite-size effects (dependence on the particle number $N$, FSE) are not significant for the results. We  have  performed this analysis 
which revealed that FSE are negligible for $N\geq 34$ (as in  Figs.~\ref{fig:covergT125} and \ref{fig:covergT31}). Therefore, the extrapolation results for the plasma composition can be considered valid, also in the thermodynamic limit. More details on the treatment of FSE are given in Sec.~\ref{sss:n-dependence}). 

We now discuss the improved extrapolation procedure that exploits the chemical model, starting from the intermediate density ($r_s=6$, Fig.~\ref{fig:convT15rs6}), where the best agreement between our HSCM-model and the FP-PIMC is observed. Indeed, the HSCM-curves reproduce the PIMC data ($N=34, 48, 64$) for pressure and energy [Figs. c), d)] within the statistical error bars for $P\geq 30$. 
Interestingly, a number $P=30$ of fourth-order propagators corresponds to an effective temperature $3 \times P \times 15,640$K$\approx 1.5\cdot 10^6$K which is close to the value $2\cdot 10^6$ K that was found sufficient to accurately reproduce the properties of an isolated hydrogen molecule by Militzer {\it et al.}~\cite{Militzer_2001}. 

Next, consider a higher density corresponding to $r_s=4$, cf. Fig.~\ref{fig:convT15rs4}. Here the degeneracy is significantly increased leading to a rapid increase of the FSP with both $P$ and $N$, and we have to restrict our FP-PIMC simulations to $N=14 (20)$. This situation makes an accurate extrapolation to $P\to \infty$ and to the thermodynamic limit very complicated. Here we strongly benefit from the strongly improved convergence of the species fractions in the chemical model. In fact, the HSCM data experience  a $1/P$-dependence similar to the scaling of the FP-PIMC data (see the black dashed line obtained by a linear extrapolation). Even though the slopes are different, in both cases the same $P\rightarrow \infty$ limit (shown by the solid horizontal sienna line) is approached.

Finally, for the lowest density case, $r_s=14$, cf.~Fig.~\ref{fig:convT15rs14} a) and b), we observe some fluctuations in the fraction of bound states with the particle number for $34\leq N\leq 64$. This is related with the slow convergence of thermodynamic averages at low densities: one has to sample a significantly larger number of configurations, giving rise to an increase of the simulation time. Therefore, we choose the $N=34$ as a reference, for $r_s=6$ and $r_s=14$.  
However, we observe that the $1/P$-slope predicted by HSCM noticeably deviates from the $1/P$-scaling of the FP-PIMC data (dashed black line). The observed discrepancy in the energy (pressure) is about $1\%$ ($3\%$), and indicates that this method is not applicable in the present case, without substantial  improvement of the HSCM.

\begin{figure}[]
\hspace{-0.9cm}
\includegraphics[width=0.51\textwidth]{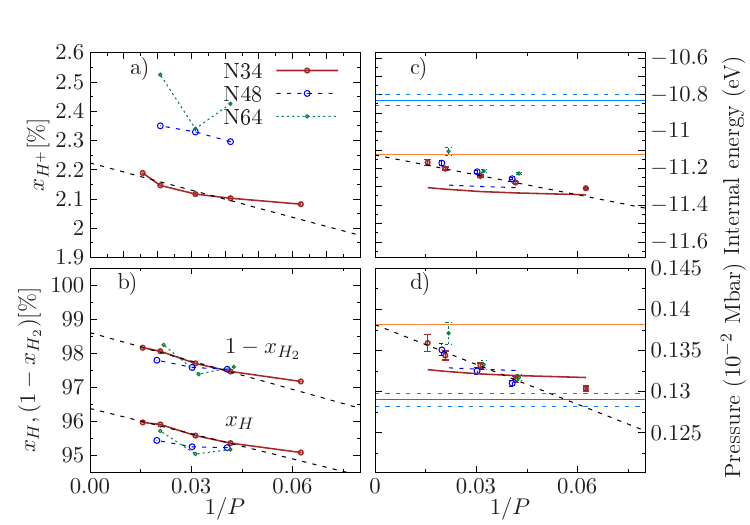}
\vspace{-0.5cm} 
\caption{Same as in Fig.~\ref{fig:convT15rs6}, but for $r_s=14$, $T=15, 640$~K and  $N=20,34,48,64$. 
The chemical model (solid brown ($N=34$) and dashed blue ($N=48$)  curves) predict different $1/P$-slopes in the FP-PIMC data, therefore, we use the latter to extract the $P\to \infty$ limit (horizontal sienna line). 
}
\label{fig:convT15rs14}
\end{figure}

Finally, we compare our extrapolated results for pressure and internal energy to the RPIMC data~\cite{Hu2011} which are shown in Figs.~\ref{fig:convT15rs6}--\ref{fig:convT15rs14} by the horizontal blue lines. While for $r_s=6$ both simulations  agree within the error-bars, cf. Fig.~\ref{fig:convT15rs6}, for $r_s=14$ and $r_s=4$ systematic deviations are observed. A more systematic analysis will be performed below for the pressure and energy isotherms.

\subsubsection{Finite size effects}\label{sss:n-dependence}

\begin{figure}[]
\hspace{-0.5cm}
\includegraphics[width=0.49\textwidth]{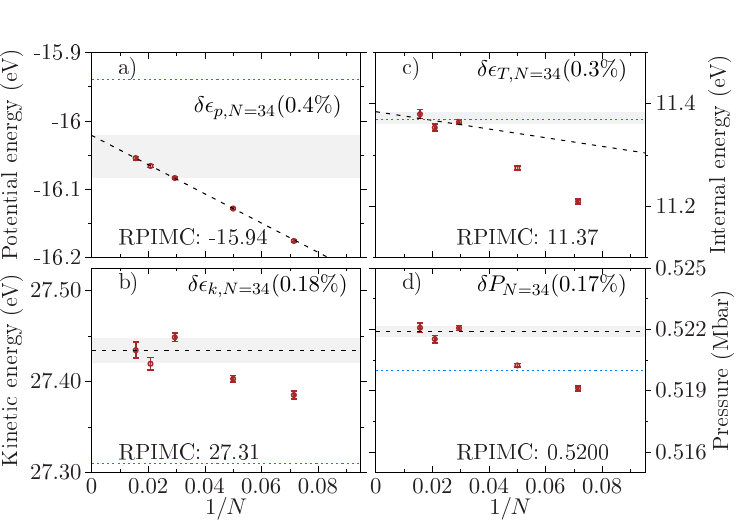}
\vspace{-0.2cm} 
\caption{Convergence of the thermodynamic quantities with the system size $N=14\,, 20\,, 34\,, 48\,, 64$ for $r_s=4$ and $T=95\,250$~K and $P=4$. Dashed lines: linear extrapolation to the thermodynamic limit. The shaded area is the confidence interval for the mean thermodynamic value $\Avr{\hat O}$: $O(\infty)- \Delta O_N$, for all simulations with $N\geq 34$. In parentheses we provide the relative deviation of the finite-size result $\Avr{O_{N=34}}$ from the asymptotic value, $\delta O_{N=34}/O(\infty)$ [$\%$]. Horizontal dotted blue lines:  RPIMC data~\cite{Hu2011}.
}
\label{fig:DiffNrs4T95}
\end{figure}

%\textcolor{red}{we need to give the magnitude of the finite size errors and demonstrate the accuracy of our simulation data. A comparison to RPIMC comes in the next sections. Move the blue text.}

\begin{figure}[]
\hspace{-0.5cm}
\includegraphics[width=0.49\textwidth]{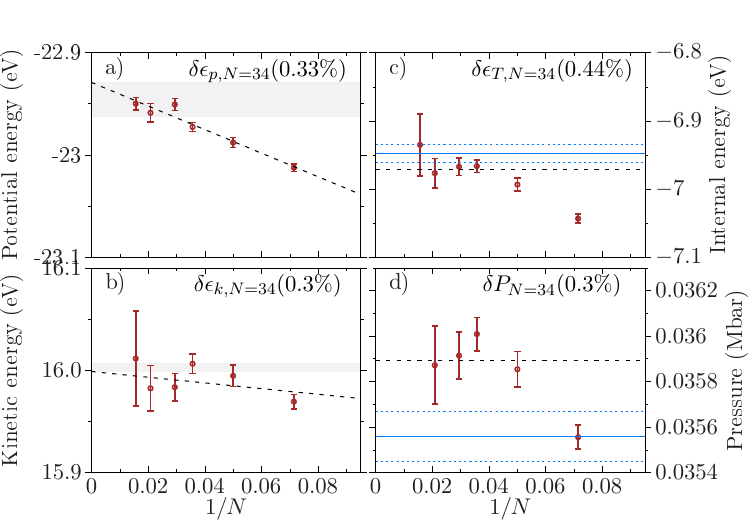}
\vspace{-0.2cm} 
\caption{Same as in Fig.~\ref{fig:DiffNrs4T95}, but for $r_s=6$ and $T=31,250$~K. The finite-size effects, $14\leq N \leq 64$, are estimated using $P=12$. The convergent result is obtained via $1/P$-extrapolation, as explained in Fig.~\ref{fig:covergT31}. 
}
\label{fig:DiffNrs6T31}
\end{figure}
In this section we analyze the influence of the finite size effects (FSE). The convergence of the FP-PIMC results strongly depends on the complexity of the FSP and, therefore, in some regions of the temperature-density plane with $n \lambda^3 \gtrsim 3$ we have to restrict the simulations to $N \leq 34$. Nevertheless, by the inclusion of periodic boundary conditions via the Ewald summation~\cite{Allen1987,Fraser1996} or the Yakub procedure \cite{yakub_ltp_05}, the $N$-dependence of the thermodynamic observables 
is substantially reduced.
This is illustrated below for two relevant cases:
the first is that of a partially ionized plasma and the second corresponds to a situation where atoms and molecules dominate.  To study the FSE we performed simulations with $N=14, 20, 34, 48, 64$ ions.
%\textcolor{red}{what value of $P$ has been used?}

For the first case we chose $r_s=4$ and $T=95, 250$K, where only few bound states are present. The four main thermodynamic functions are plotted in Fig.~\ref{fig:DiffNrs4T95} and exhibit an almost  linear $1/N$-scaling.  
%The shaded area specifies the interval centered at the extrapolated value, $\Avr{O}_{\infty}$, which is expected to contain all finite-size results $\Avr{O}_N$ for $N \geq 34$.  
The relative deviation of the reference system size ($N=34$) from the TDL in percent is indicated in each panel and provides a quantitative estimate of possible finite-size errors. This also applies to the values reported in Appendix~\ref{eos-table}, including the special cases when, due to the FSP, the simulations were restricted to $N\leq 20$. 
 
Several important conclusions can be drawn. The smallest (and the largest) FSE are observed in the kinetic $\epsilon_k$ (and the potential $\epsilon_p$) energy contributions. The kinetic energy  estimator~(\ref{Ek_final}) has a factor $3-4$ larger statistical error compared to $\epsilon_p$, but it is much less influenced by the FSE than the potential energy. In particular, for $N=20$ the estimated kinetic energy deviates from the TDL by $\delta \epsilon_{k,N=20}\sim 0.2\%$, while the deviations in the potential energy reach up to $\delta\epsilon_{p,N=20}\sim 1\%$.
Since the internal energy~(\ref{ep}) and the pressure~(\ref{Pvir}) contain both quantities (with different weights), the related FSE can be reduced significantly, once they are removed from the potential energy~\cite{dornheim_prl16,filinov_cpp_21}.

The second case, corresponding to $T=31, 250$K and $r_s=6$, cf. Fig.~\ref{fig:DiffNrs6T31}, exhibits very similar scaling relations. Note, that the FSE are practically absent in the kinetic energy (panel b) and pressure (panel d), even for the smallest system size $N=20$. For the lower temperature $T=15, 640$K (not shown) this occurs even for $N=14$. In this regime the plasma composition is dominated by  bound states (see below), and an almost ideal neutral gas behavior is expected at $r_s \geq 6$ with only a weak dependence on the system size. 

After having analyzed the convergence of our FP-PIMC simulations with respect to $P$ and $N$ we now turn to a discussion of the thermodynamic properties.

\subsection{EOS at high temperatures and low densities: the non-degenerate case}\label{DHmodel}

\begin{figure}[t]
\hspace{-.7cm}
\includegraphics[width=0.515\textwidth]{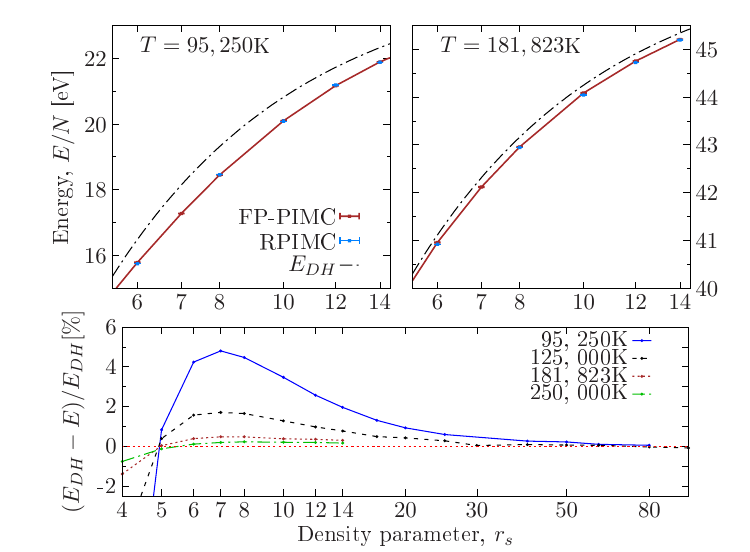}
\vspace{-0.5cm}
\caption{{\it Upper panels:} Internal energy per particle versus the density parameter, 
%$4 \leq r_s \leq 100$, 
%($1.96\cdot 10^{-3}\text{g/cm}^3 \leq \rho \leq 4.3 \cdot 10^{-2}$g/cm$^3$) 
for $T=95,250$K and $181,823$K. The FP-PIMC ($N=64$) is compared to RPIMC EOS~\cite{Hu2011} and the Debye-Hückel limit (DH)~(\ref{E_DH}), $E_{\text{DH}}$ (dashed-dotted line). {\it Lower panel:} Relative deviations of the FP-PIMC energy from $E_{\text{DH}}$ for four temperatures. }
\label{fig:DH-modelE}
\end{figure}

In this section we analyze the EOS at high temperatures, $T\geq 95,250$K, and low densities, $r_s\geq 5$ ($\rho_H\leq 2.15 \cdot 10^{-2}$g/cm$^3$), where the plasma is non-degenerate and strongly or even completely ionized. Isotherms of total energy and pressure are shown in Figs.~\ref{fig:DH-modelE} and~\ref{fig:DH-modelP} and display the expected monotonic increase with $r_s$. For $k_BT> 1Ry$ (right figures), in the low-density limit, the pressure approaches the classical ideal gas result, $p_{cl}=(n_e+n_i)k_BT$, and the energy, $2 \frac{3}{2}k_BT$. When the density increases, interaction effects grow, and the leading correction to the ideal gas behavior is given by the Debye-Hückel limit (DH~\cite{DHmodel}): 
\begin{eqnarray}
&&\beta p_{\text{DH}}=\beta p_{\rm id}-\frac{\kappa^3}{24 \pi}, \label{P_DH}\\ 
&&E_{\text{DH}}=E_{\rm id}-k_B T \frac{\kappa^3}{8 \pi n} \label{E_DH},
\end{eqnarray}
where the inverse Debye length, $\kappa^2=4 \pi n e^2/k_B T$, is defined by the full density, $n=n_i+n_e$, and we introduced
$ \beta p_{\rm id}= n_i+ \beta p^{\rm id}_e\,,
E_{\rm id} =\frac{3}{2} k_B T+\frac{2}{3} \frac{p^{\rm id}_e}{n_e}\,,$
where $p^{\rm id}_e$ is the pressure of an ideal Fermi gas.
\begin{figure}[]
\hspace{-.5cm}
\includegraphics[width=0.5\textwidth]{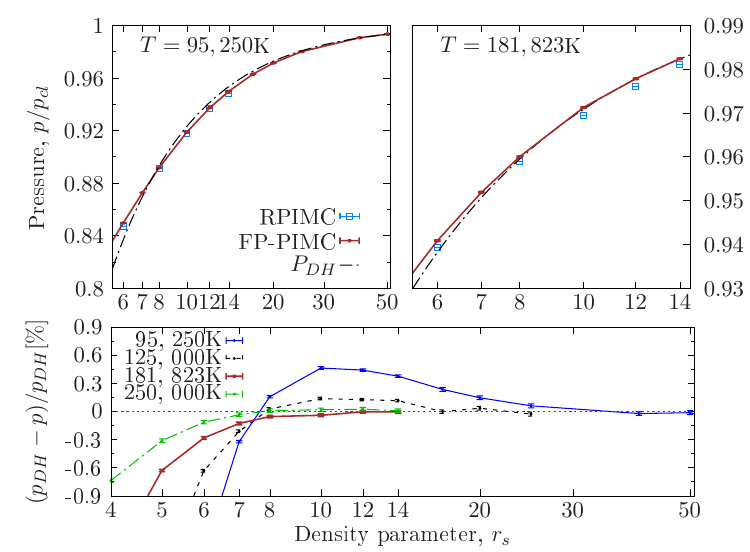}
\caption{Same as Fig.~\ref{fig:DH-modelE}, but for the pressure isotherms plotted in units of $p_{cl}=(n_e+n_i)k_BT$. 
}
\label{fig:DH-modelP}
\end{figure}
Extending the FP-PIMC simulations to low densities, $r_s=100$, we can establish the validity range of the DH limit: it is generally very accurate for $r_s \gtrsim 20$ and, when the temperature increases, the density range grows towards smaller $r_s$, cf. lower panels of Figs.~\ref{fig:DH-modelE} and~\ref{fig:DH-modelP}. 
In contrast, when the density increases, $r_s\lesssim 20$, the DH approximation underestimates the (negative) correlations in the plasma, and it entirely misses bound states.

Finally, we compare the FP-PIMC results to the RPIMC EOS~\cite{Hu2011} shown by the blue symbols in the top parts of Figs.~\ref{fig:DH-modelE} and \ref{fig:DH-modelP}. For the energy isotherms, we observe perfect agreement in the entire density range where RPIMC data are available ($r_s \le 14$). At the same time, RPIMC slightly underestimates the pressure.
This good agreement could be expected, as in the density  range of $5\leq r_s\leq 14$ the electron degeneracy factor
varies between $4 \leq \theta \leq 61$, i.e. the electrons are non-degenerate, and the fixed-node approximation does not have a significant impact on the simulations.

%\subsection{Results for high temperature and high density}\label{highdens}
%
\begin{figure*}[t]
\hspace{-1.2cm}
\includegraphics[width=0.74\textwidth]{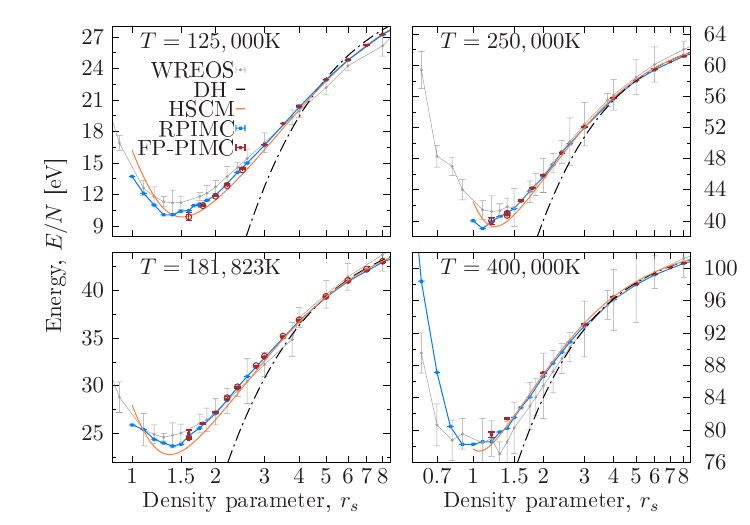}
\vspace{-.4cm}
\caption{Isotherms of internal energy for four temperatures $T=125,000 - 400,000$K. Red symbols: FP-PIMC data for $N=34(64)$. Blue symbols:  RPIMC, grey symbols with error bars: WREOS; dash-dotted line: Debye-Hückel limit (DH); orange lines: chemical model (``HSCM''). 
}
\label{fig:ET125}
\end{figure*}
\subsection{EOS at moderate and high densities: the  case of degenerate electrons}\label{highdens}
We now turn to higher densities, $1.5\leq r_s \leq 8$, corresponding to a hydrogen mass density $5 \cdot 10^{-3}\text{g/cm}^3 \leq \rho_H\leq 0.80\,$g/cm$^3$, where electron degeneracy effects become important. The results are presented in Figs.~\ref{fig:ET125}--\ref{fig:PressEnergT62}. 
%and \ref{fig:PressEnergT95}.

% \textcolor{red}{First discuss the FP-PIMC results and bring comparison to DH, RPIMC and the others later.} 

We start by considering the case of high temperatures, $T\ge 125,000$ K where the plasma is (almost) fully ionized. Figure~\ref{fig:ET125} shows four energy isotherms and compares our FP-PIMC results to alternative models that were introduced in Sec.~\ref{ss:hydrogen-results}.
The overall behavior of the isotherms is well known: from the low-density limit the energy monotonically decreases, due to an increase of (negative) Coulomb correlations. Upon further density increase growth of Coulomb correlations competes with a more rapid increase of quantum kinetic energy resulting in an energy minimum around $r_s = 1\dots 1.5$, after which the energy increases steeply. Our FP-PIMC simulations allow us to accurately determine the total energy isotherms (cf. red symbols in Fig.~\ref{fig:ET125}) and to come close to the energy minimum: we reach $r_s=1.6$, for $T=125,000$K and $r_s=1.2$, for $T\ge 250,000$K.
For $r_s < 1.2$ FP-PIMC simulations are not feasible due to the FSP:  the electron degeneracy parameter reaches  $\theta=0.53$, whereas the average sign, Eq.~(\ref{def_Sign}), drops below $2 \cdot 10^{-3}$. 

The analysis of the energy isotherms is extended to lower temperatures in Figs.~\ref{fig:PressEnergT95} and \ref{fig:PressEnergT62} where we, in addition, include also the equation of state. For $T=95,250$ K our simulations are possible up to $r_s=2$. Interestingly, while we cannot access the energy minimum (which is around $r_s=1.5$), we completely resolve the pressure minimum which occurs at significantly lower densities (around $r_s\approx 2.5$). The same is observed for $T=62,500$~K, cf. Fig.~\ref{fig:PressEnergT62}, where the error bars are still reasonably small. For $T=31,250$~K we reach $r_s=4$ which is very close to the minimum, but at least another data point at higher density would be needed to give a conclusive answer.
Thus we can provide first-principle data for the location and depth of the pressure minimum, for $T\ge 62,500$~K. Since this minimum arises from a competition of a variety of physical effects (see above) which are difficult to capture simultaneously in simpler models, our data  constitute highly valuable benchmarks for other models.

Due to the limitations of the FP-PIMC simulations by the FSP, it is interesting to explore how accurate 
the chemical model (HSCM) is that was introduced and applied in Sec.~\ref{sss:composition-vs-p}, and whether it is suitable to provide an extension of the isotherms to smaller $r_s$. It turns out that the HSCM model
 (solid orange lines) is particularly well adopted to the energy isotherms and is accurate for all densities in the very broad range $r_s \geq 3$. Also, the HSCM model seems to qualitatively capture the behavior of the total energy around its minimum,  
 up to $r_s\sim 1$, for $T\ge 125,000$~K and up to $r_s\sim 1.2$, for $T\le 125,000$~K. On the other hand, the present HSCM model is much less accurate for the pressure, see left parts of Figs.~\ref{fig:PressEnergT95}--\ref{fig:EOST15}.
 
We now turn to the comparison with the results from other models, cf. Sec.~\ref{ss:hydrogen-results}. Consider first the comparison with the RPIMC results. For all FP-PIMC data shown in Figs.~\ref{fig:ET125}--\ref{fig:PressEnergT62} we observe agreement with RPIMC, within the statistical errors.

Next, we compare to WREOS -- the wide-range EOS by Wang {\it et al.}~\cite{Wang}. 
The agreement of the energies for $T\ge 125,000$~K is very reasonable within the provided error-bars but, apparently, the energy minimum is underestimated. Similar trends are observed for the EOS and for lower temperatures
and become even more pronounced for temperatures $30, 000\text{K}\leq T\leq 100, 000$K, cf. Fig.~\ref{fig:PressEnergT62}. Due to the large error bars, we did not include the data for $T=95,250$~K into Fig.~\ref{fig:PressEnergT95}.    
Note that the DFT data of Ref.~~\cite{Wang2013} for the energy contain an unknown constant. To plot the data  in Fig.~\ref{fig:PressEnergT62}, the single molecule ground-state energy [-15.502 eV] was subtracted. 
%in the DFT data by Wang {\it et al.}, but the comparison can be not fully conclusive as the DFT-pressure deviates significantly from the RPIMC and the FP-PIMC data, predicting higher values.

%The extraction of the absolute value of the internal energy from the DFT simulations is another problem which is not straightforward. The DFT data need to be normalized either by a characteristic energy of an isolated bound complex or using a theoretically known asymptotic behavior in the low-density limit~\cite{Becker_2014}. 

%
\begin{figure}
\hspace{-.6cm}
\includegraphics[width=0.52\textwidth]{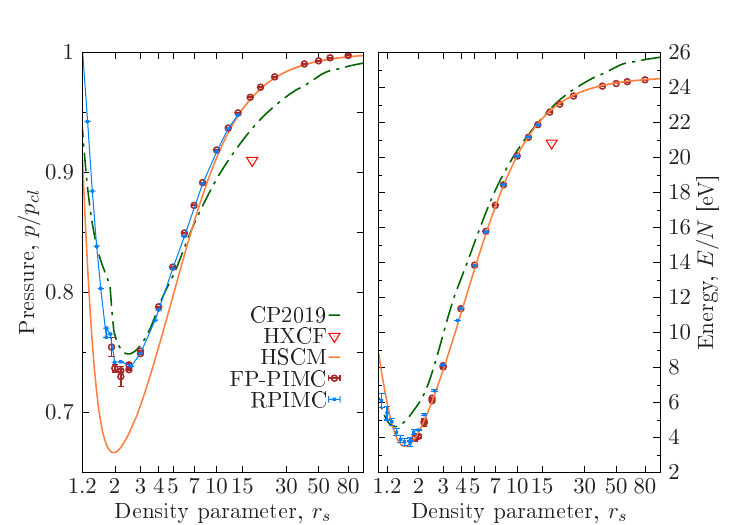}
\caption{Isotherms of pressure (left -- in units of $p_{cl}=2nk_BT$) and internal energy (right), for $T=95,250$K. Small red symbols: FP-PIMC data for $N=34$  and $N=20$ (at $r_s  \leq 2$); Blue symbols:  RPIMC; dash-dotted green line: CP2019; Red triangle at $r_s=17.53$: HXCF; Orange line: ``HSCM''; 
%gray line: ``CM'', 
cf. App.~\ref{ss:chem-model}.}
\label{fig:PressEnergT95}
\end{figure}

Consider now the comparison to the low-density result ($r_s=17.53$) from the ``HXCF'' data by Mihaylov {\it et al.}~\cite{Karasiev_2021}, cf. Sec.~\ref{ss:hydrogen-results}. 
We observe very good agreement for the energy (cf. the red triangle in Fig.~\ref{fig:PressEnergT62}) and a minor discrepancy in the pressure. Further,  we observe the general trend that ``HXCF'' starts to systematically deviate from the FP-PIMC data with increasing temperature. While we find a nearly perfect agreement at $31, 250$K, significant deviations appear at $95, 250$K, see Fig.~\ref{fig:PressEnergT95}. 

We now turn to the isotherms labeled ``CP2019'', by Chabrier~{\it et al.}~\cite{Chabrier_2019}, cf. Sec.~\ref{ss:hydrogen-results}, which are included in Figs.~\ref{fig:PressEnergT95} and \ref{fig:PressEnergT62}.
These data extend to low densities allowing for a comparison in the range from $r_s\approx 3$ to $r_s=100$.
Overall, the agreement with the FP-PIMC curves is good, with the pressure data being more accurate whereas the energies are systematically too high.
\begin{figure*}[]
\hspace{-.6cm}
\includegraphics[width=0.51\textwidth]{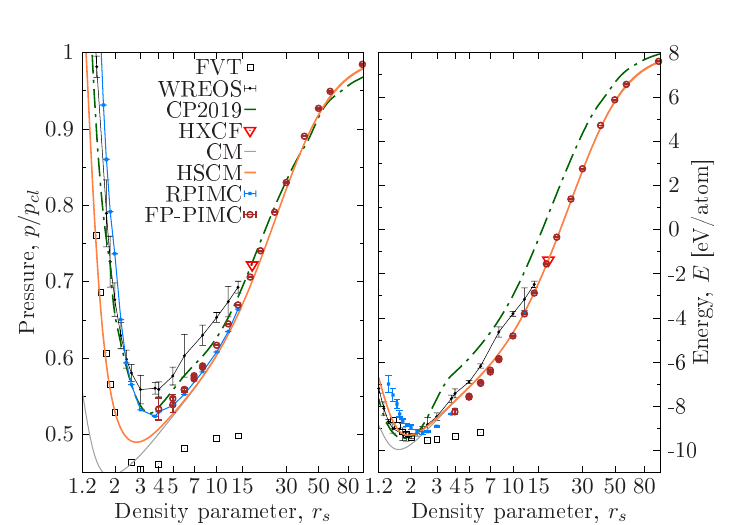}
\includegraphics[width=0.51\textwidth]{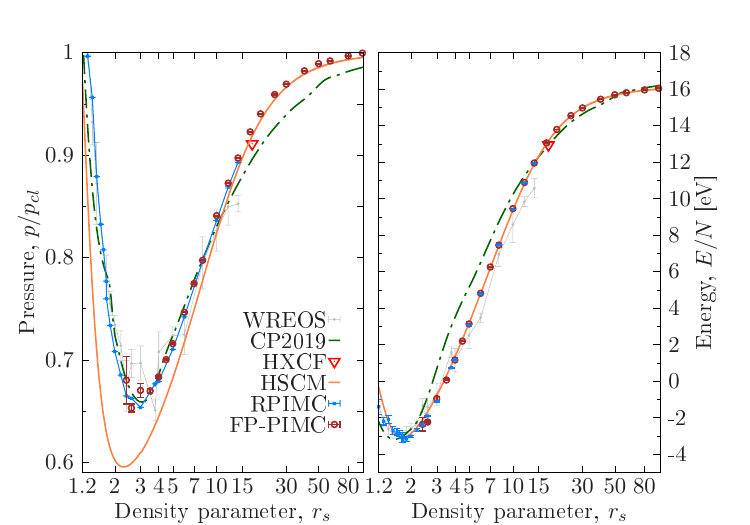}
\caption{Same as in Fig.~\ref{fig:PressEnergT95}, but for $T=31,250$K (left) and $T=62, 500$K (right). Open red circles: FP-PIMC data; solid blue lines with symbols: RPIMC; grey lines with error bars: WREOS; open squares:  ``FVT'': green dash-dotted lines: ``CP2019''; red triangles at $r_s=17.53$: ``HXCF''; orange lines: ``HSCM'', grey lines in the left figure: ``CM'', cf. App.~\ref{ss:chem-model}.
}
\label{fig:PressEnergT62}
\end{figure*}

The fluid variational theory (FVT) by Juranek {\it et al.}~\cite{Juranek2002} is included for the 31kK isotherms. We observe that, both, the total energies and pressure are substantially too low, where a comparison is possible, i.e. for $r_s \in [4, 15]$. This is apparently caused by the contributions of unbound electrons, so the improved FVT+ model of Holst \textit{et al.}~\cite{Redmer2006,Holst2007} is expected to be more accurate.

\subsection{EOS in the atomic and molecular regime}\label{lowT}
\begin{figure*}[t]
\hspace{-.6cm}
\includegraphics[width=0.74\textwidth]{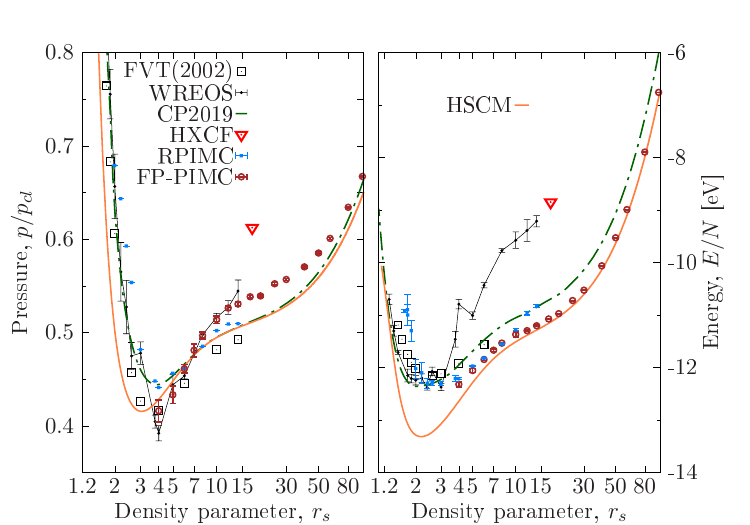}
\caption{Same as Fig.~\ref{fig:PressEnergT62}, but for $T=15, 640$~K. Open red circles: FP-PIMC with $P=72\dots 96$ and $N=34$, for $r_s\geq 4.5$ [$N=14$, for $r_s\geq 3.7$];   blue symbols: RPIMC; 
%grey lines with dots: H-{REOS}.3 ($T=15,000$~K); 
black lines with error-bars: WREOS; open squares: ``FVT''; dash-dotted green lines: ``CP2019'';  red triangle at $r_s=17.53$: ``HXCF''; orange lines: ``HSCM'', cf. App.~\ref{ss:chem-model}. } 
\label{fig:EOST15}
\end{figure*}
\begin{figure*}[]
\hspace{-0.7cm}
\includegraphics[width=0.74\textwidth]{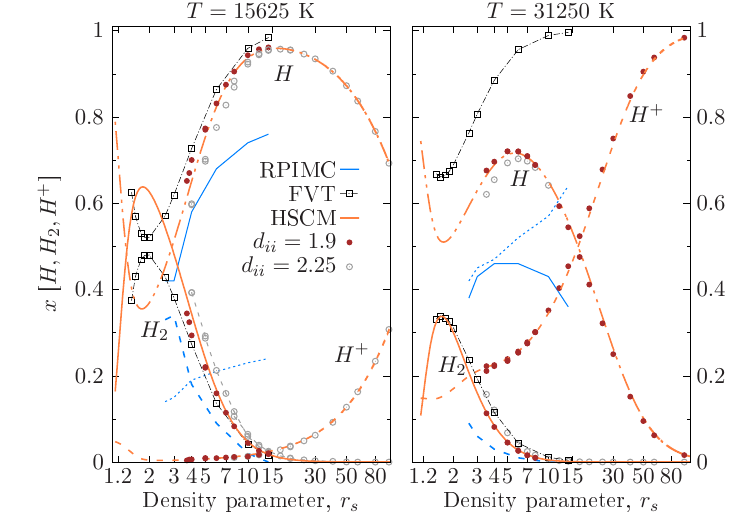}
\vspace{-0.4cm} 
\caption{Fractions of molecules, atoms and free protons, for two isotherms,   $T=15,640$~K (left) and $T=31,250$~K (right). FP-PIMC results are plotted for $d^{cr}_H = 1.9 a_B$ (brown solid dots) [$d^{cr}_H = 2.25 a_B$, open gray circles], for details, see text. Blue lines: RPIMC data for $d^{cr}_H = 1.9 a_B$; open squares:  atom and molecule fractions from the FVT. Orange lines: ``HSCM'', cf. Sec.~\ref{ss:chem-model}. }
\label{fig:Nfrac}
\end{figure*}

Now we turn to the lowest temperature in the considered range, $T=15, 640$~K, where the thermodynamic functions are dominated by the contributions of neutrals. The isotherms of pressure and energy follow the trends discussed for the higher temperatures, cf. Fig.~\ref{fig:EOST15}. Again, our simulations are severely hampered by the FSP -- here we are limited to $r_s \ge 3.5$. We  use $N=14$ ($N=20$) particles for $r_s \in [3.5,5]$, and $N=34$, for larger $r_s$. Nevertheless convergence with respect to $P$ and $N$ has been achieved, but the error bars are increasing towards lower $r_s$.
At the same time, we observe very good agreement of our chemical model (HSCM) with the simulations, cf. the energy isotherms in Fig.~\ref{fig:EOST15}. This indicates that the cluster analysis used to extract ``fractions'' of atoms and molecules from the FP-PIMC simulations is consistent (see below).

While we had observed very good agreement of RPIMC with our simulations, for $T\ge 31,250$~K before, in the present case the deviations are significantly larger, and it is interesting to analyze them in more detail. Consider first, the energy. Here we observe excellent agreement for $r_s\in [5,10]$. For $r_s =4$ the RPIMC energy is too high by about $1.5\%$ whereas for larger $r_s$ deviations exceed $3\%$. Let us now turn to the pressure isotherms. Here both simulations agree for $r_s=6$, but the FP-PIMC shows a much stronger slope in this point. Consequently, the RPIMC pressure is too high, for smaller $r_s$ (up to $8\%$) and too low for large $r_s$ (up to $5\%$).

%First, the RPIMC pressure and energy isotherms are rather close to our data, for $6 \lesssim r_s \lesssim 12$). However, deviations are observed outside this range. For high densities, $r_s\in [3.5, 6]$ both, RPIMC pressure and energy are significantly above our results with the deviations reaching about $5\%$, for the energy, and $10\%$, for the pressure. Even more surprising are deviations observed for low densities, $r_s \gtrsim 10$: here the RPIMC pressure falls below our data and exhibits an unexpected saturation. In contrast, the total energy strongly deviates towards positive values.

%Finally, even in the ab initio RPIMC-data for $10\leq r_s\leq 14$ we observe some strange saturation in the compressibility factor, $p/p_{cl}$, and, simultaneously, the energy isotherm starts to deviate more strongly both from the HSCM and the FP-PIMC. We attribute this effect to the insufficient convergence of the thermodynamic functions (statistical averaging in the RPIMC should be improved).  

A more detailed comparison of the FP-PIMC and RPIMC results is achieved by analyzing the microscopic configurations of the electron paths, in particular, their spatial extension. Even though the PIMC approach does not distinguish between bound and free electrons, an (artificial) distinction can be introduced via a cluster analysis, as demonstrated by Militzer \textit{et al.} in Ref.~\cite{Militzer2001}. They introduced a critical average proton--proton separation, $d_H^{cr}=1.9 a_B$, below which the configuration was ``counted as a molecule''. Even though this threshold value has no direct physical meaning, it allows to better compare different PIMC simulations. 
This value is a reasonable estimate for the spatial extension of a molecule, if we consider the slope of the ion-ion PDF at this temperature, see Fig.~\ref{fig:PDFT15P}.

We have used the same criterion for the molecules but use a modified criterion for the atoms, as explained below. The results for $T=15,625$~K and $T=31,250$~K are shown in Fig.~\ref{fig:Nfrac}. 
For both temperatures, we observe reasonable agreement for the molecule fractions. However, the fraction of atoms (free protons), in the FP-PIMC simulations is significantly higher (lower) than in the RPIMC data.
However, since this is the case for both temperatures, whereas the thermodynamic functions, for $T=31,250$~K, are in very good agreement, this cluster analysis does not fully explain the discrepancies, see also Sec.~\ref{s:conclusion}.

For completeness, we explain how we define the atom fraction in our FP-PIMC cluster analysis. 
%We analyze the ensemble of particle trajectories. 
For each ion trajectory ($\{\vec r_{p,j}=\vec r_j(\tau_p), \, 0 \leq \tau_p \leq  \beta \}$)  we calculate the total charge due to all electrons within a sphere of radius $R_a=3a_B$
\begin{eqnarray}
   \rho_{\text{net}}^{I, j}=\frac{1}{P} \sum\limits_{i=1}^{N_e} g_{p,i} \, \theta(\abs{\vec r_{p,i}-\vec r_{p,j}})\,,
\end{eqnarray}
averaged along the imaginary time, and
$$
\theta(x)=
\begin{cases}
1, \abs{x} \leq R_a \\
0, \;\rm{else}\,.
\end{cases}
$$
The weighting factor $g_{p,i}$ takes into account the possibility that each electron defined by the vector $\vec r_{p,i}=\vec r_i(\tau_p)$, at a given imaginary time, can be simultaneously within the radius $R_a$ of several ($N_p^{I}$) atoms. In this case its contribution on the given time slice $\tau_p$ is equally distributed between the nearby $N_p^{I}$ ions with the weight $g_{p,i}=1/N_p^{I}$.
Following this recipe we treat an ion as  ``neutral'' (belonging to an atom) or free particle depending on the accumulated net charge $\rho_{\text{net}}^{I, j}$, according to the  criterion
$$
\abs{\rho_{\text{net}}^{I, j}/e}=
\begin{cases}
\leq 0.2,\quad  \text{neutral},\quad N_a \to N_a+1\,, \\[1ex]
\geq 0.8,\quad  \text{free ion},\quad N_i\to N_i+1\,,
\end{cases}
$$
where we indicated that for the cases we increase either the number of atoms or the number free ions by one.

The final fractions of free and bound ions are obtained by statistical averaging, as in the case of other observables.
Finally, If two ``neutrals'' are found at a distance $d_{ii}\leq 1.9$ (or 2.25) $a_B$, they are counted as a molecule, as discussed above, and we update the number of molecules,  $N_m \to N_m+1$.
The corresponding particle fractions are determined by the statistical averaging similar to other thermodynamic quantities
$$
\Avr{N_{i(a,m)}}=\frac{\left\langle N_{i(a,m)} \cdot  \prod\limits_{p=1}^P \sgn_p \right\rangle}{ \left\langle \prod \limits_{p=1}^P\sgn_p\right\rangle}
$$
where $\sgn_p$ is the sign of configurations due to the Slater determinants in the fermionic partition function~(\ref{ZF2}).
In Fig.~\ref{fig:Nfrac} we plot the corresponding fractions of ions, atoms and molecules,
$$
x[H^+]=\frac{\Avr{N_i}}{N},\quad  x[H]=\frac{\Avr{N_a}}{N}, \quad x[H_2]=\frac{2\Avr{N_m}}{N}.
$$

Note that the fractions of atoms and molecules that are obtained from the described cluster analysis maybe given some physical significance if the results is reasonably independent of the chosen threshold values. To this end we varied these values. As an illustration, we included results for a typical second case for the molecules, $d_H^{cr}=2.25 a_B$, in Fig.~\ref{fig:Nfrac}. Obviously, the influence of the threshold is rather small which allows us to use these results in our chemical model, as well as for comparison with other approaches, such as the FVT, the results of which are also presented in Fig.~\ref{fig:Nfrac}. While the agreement with FP-PIMC is very good, for $T=15,625$~K, the model fails for $T=31,250$~K because, in the latter case, already a significant fraction of free particles ($20\%\dots 30\%$) is present.

Finally, for an additional comparison between RPIMC and FP-PIMC, in Fig.~\ref{fig:EkinEpot} we plot the kinetic and the potential energy isotherms, for three temperatures. For $T=31,250$~K and $T=62,500$~K where the total energies of RPIMC and FP-PIMC agreed within statistical errors, similar agreement is observed for the potential energy. On the other hand, we observe noticeable deviations of the kinetic energies. This shows that the kinetic energy is an observable that is very sensitive to details and, possibly errors, of the simulation procedure, whereas in the total energy deviations are reduced due to possible error compensations. This observation is confirmed for $T=15,640$~K. Here, for $r_s \gtrsim 6$, we observe significant deviations of both potential and kinetic energy which have opposite sign.\\

Finally, let us briefly summarize the comparison with the other models that are also included in Fig.~\ref{fig:EOST15}. While the DFT-based ``WREOS'' data showed reasonable agreement with FP-PIMC, for $T\ge 31,250$~K, here we observe stronger deviations. For the pressure, the agreement is rather good, except for $r_s=4$ and $r_s=14$. On the other hand there are significant deviations for the internal energy which rapidly increase with $r_s$. Similar large deviations are observed for the ``HXCF'' data point for the energy at $r_s=17.53$. However, ``HXCF'' also strongly disagrees for the pressure.
The EOS ``CP2019'' exhibits similar trends as for $T\ge 31,250$~K. Here the largest deviations (of the order of several percent) for the pressure are observed around the minimum (positive deviations) and in the range $10 \lesssim r_s \lesssim 50$ (too low values). For the energy there appears to be an almost constant positive shift compared to FP-PIMC.
Finally, the ``FVT''-curve for the pressure (energy) proceeds close to the ``CP2019'' isotherm, for $r_s$-values larger than $6$ ($3$) and, hence, exhibits similar deviations from FP-PIMC. 
%Finally, the ``H-REOS.3'' curves (available for $T=15,000$~K) have the advantage that they also extend to very low densities, $r_s=100$. They are of similar accuracy as the ``CP2019'' data when the temperature correction is taken into account.\textcolor{red}{we need to verify this!}

\begin{figure}[]
%\hspace{-.5cm}
\includegraphics[width=0.5\textwidth]{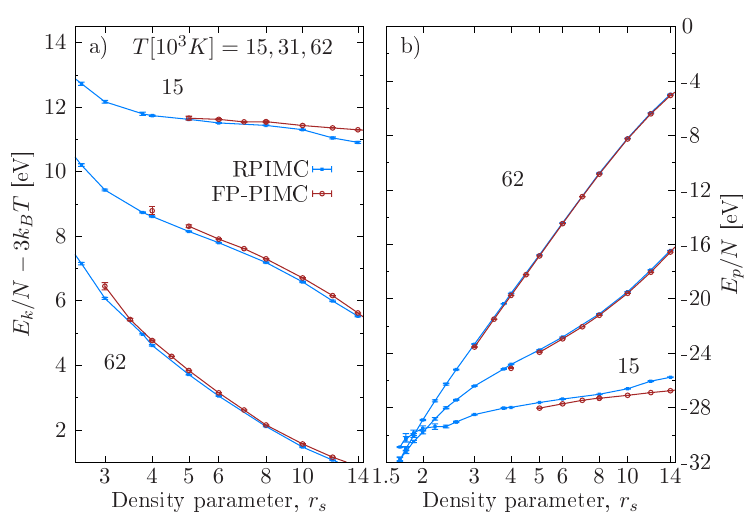}
\vspace{-.5cm}
\caption{Isotherms of (a) kinetic energy and (b) potential energy for three temperatures indicated in the figure. FP-PIMC and RPIMC data are compared. }
\label{fig:EkinEpot}
\end{figure}

\section{Conclusions and outlook} \label{s:conclusion}
With the upcoming new experimental facilities, in particular the 
colliding planar shocks platform at the NIF \cite{nif-eos_23} and the
FAIR facility at GSI Darmstadt \cite{tahir_cpp19}, high precision thermodynamic data for highly compressed matter will be available. This poses new challenges to theory and simulations. While presently a large variety of competing models and simulation approaches are being used  the predictions of which often deviate significantly from one another, no hard experimental benchmarks have been available or the experimental error bars are too large for a discrimination. On the other hand, existing restricted path integral Monte Carlo simulations, which are expected to be the most accurate approach, are computationally expensive, and no independent test of their accuracy or range of applicability has been available.
 
In this article we have presented extensive independent fermionic PIMC simulation results for dense partially ionized hydrogen (for the present parameters these apply also to deuterium). These simulations avoid the fixed node approximation and are thus free of systematic errors. 
Therefore, our simulations are well capable to serve as benchmarks for RPIMC and other approaches.
At high densities, the present FP-PIMC simulations are severely hampered by the fermion sign problem which restricts simulations to moderate degeneracy of the electrons -- here we had to limit ourselves to temperatures above $15,000$~K and densities in the range of $r_s \in [2, 100]$. We have presented details of our fermionic propagator PIMC approach and demonstrated in detail convergence with respect to the simulation size $N$ and the number $P$ of fourth-order propagators. The results for the equation of the state, energy contributions and pair distributions should be valuable for benchmarking and possibly improving alternative methods. Even though PIMC simulations are performed in the physical picture where no artificial distinction between ``free'' and ``bound'' electrons is necessary, we performed a cluster analysis of the spatial extension of the electrons and approximately extracted the degree of ionization and dissociation. These results are also valuable reference for other methods.
In contrast to chemical models, 
which are hampered by an unavoidable inconsistency in the treatment of the interaction between charged particles, neutral particles and between charges and neutrals, PIMC treats all interactions and exchange effects selfconsistently. The results only depend on two parameters -- the  relevant average ``spatial extension'' of an atom and of a molecule, respectively, but the influence of this choice is small and easy to quantify, as was shown in Fig.~\ref{fig:Nfrac}.

Let us summarize our comparison with the available RPIMC  data by Militzer \textit{et al.} for partially ionized hydrogen. Based on earlier comparisons against first-principle CPIMC and PB-PIMC simulations for the uniform electron gas (UEG)\cite{schoof_prl15, groth_prb16, dornheim_physrep_18} -- a much simpler system without bound states -- which revealed errors on the order of $10\%$ for high densities, similar deviations could be expected. Presently, no CPIMC simulations for hydrogen are available, due to the increase of the FSP for two-component simulations. Thus, the accuracy of existing hydrogen data for high densities and strong degeneracy on the order of $\theta \lesssim 0.5$ has to be left open.

On the other hand, the present FP-PIMC simulations provide the first accurate fermionic PIMC simulations in the complementary range of $\theta \gtrsim 0.5$, translating into density parameters $r_s \gtrsim 4$, for $T\ge 15,640$~K and smaller values, for higher temperatures. As we have demonstrated, this is sufficient to reach the minimum of the pressure isotherms, and come close to the energy minimum.
%
%where the structure of the nodal surface~\cite{Ceperley1991} has no-effect on their stability, as in the case of deuterium or hydrogen~\cite{Militzer2000}. In our recent analysis, the FP-PIMC results for the interaction energy and the exchange correlation free-energy~\cite{filinov_cpp_21}  has been benchmarked  against the RPIMC by Brown {\it et al.}~\cite{Brown_2014} and the CPIMC/PBPIMC by Dornheim {\it et al. }~\cite{dornheim_prl16}. The role of the exchange-correlation effects in the PDF functions for the spin-like electrons has been explored and revealed that the XC hole (due to the Pauli blocking) is {\it underestimated} in the fixed-node approximation for the N-body density matrix~\cite{fermion_nodes}. This in its turn, has lead to a noticeable effect in all thermodynamic observables in the UEG case~\cite{filinov_cpp_21}.
%
    Summarizing, we conclude that the comparison 
    for partially ionized hydrogen reveals overall very good agreement between RPIMC and FP-PIMC in the entire density range where both data sets are available, for temperatures as low as $T=31,250$~K. We interpret this as a strong independent confirmation of the extended first-principle simulations reported from the RPIMC method. 
    
A more detailed comparison can be conducted from the
EOS,  Tab.~\ref{tab:data} (see Appendix). Depending on the density-temperature range we resolve some systematic deviations, both in the pressure and energy, which are well above the statistical errors. In particular, for $T > 62kK$ the deviations stay below 1\%. The FP-PIMC EOS contains results for
two system sizes, $N = 34(64)$, and allows us to estimate the influence of FSE which are small. Therefore, the main reason for the
observed discrepancies, is due to
the fixed-node approximation.
Consider, in particular, the lowest temperature
isotherm $T = 15, 640$ K, cf. Fig.~\ref{fig:EOST15}, where in the
range of the minimum, deviations of up to 8\% are
predicted (the RPIMC data are too high). 
    As we showed in Fig.~\ref{fig:Nfrac}, the RPIMC results significantly underestimate (overestimates) the fraction of bound complexes (free particles), in comparison to the FP-PIMC results. 
%    This seems to be consistent with an up-shift of the RPIMC energy isotherms compared to FP-PIMC observed in Fig.~\ref{fig:EOST15}. 
Also, the deviations of the RPIMC results for pressure and energy increase for lower densities, $r_s\gtrsim 12$,  where the RPIMC data are too high by several percent. This is unexpected since there is no FSP at these parameters. Therefore, this discrepancy is, most likely, not related to the fixed node approximation but could rather reflect a sampling problem. Here, improved RPIMC data that also extend to $r_s=100$ would be desirable.

Finally, let us briefly summarize the results of the comparison to other methods as introduced in Sec.~\ref{ss:hydrogen-results} in the parameter range where FP-PIMC results have been reported. First, the wide range equation of state of Chabrier \textit{et al.} (``CP2019'') 
%and Becker \textit{et al.} (``H-REOS.3'') 
exhibits overall a good accuracy capturing the main trends. The largest deviations for pressure and energy are on the order of several percent, in the range of the minimum of the isotherms, but there are also systematic deviations for low densities. 
On the other hand, the DFT-based wide-range data sets of Wang \textit{et al}. (``WREOS'') and Mihaylov \textit{et al.} (``HXCF'') show a different behavior. WREOS is rather accurate for $T\gtrsim 250,000$~K, but exhibits increasing discrepancies when the temperature is lowered. For the lowest temperature, $T=15,640$~K, the pressure isotherm is very accurate, however, large deviations are observed for the energy. For the comparison with HXCF we avoided parameters that are designated in Ref.~\cite{Karasiev_2021} as ``interpolation''. This left us with the comparison for the density $r_s=17.53$ where the simulation method is clear. There we observed very good agreement for $T=31,250$~K and $T=62,500$~K, but significant discrepancies for $T=95;250$~K and even larger
deviations for $T=15,640$~K. 
Based on the comparison with a variety of DFT-MD simulations (not shown) our data should be able to discriminate between different exchange-correlation functionals. LDA-type functionals are clearly not sufficient, even though the finite temperature version (GDSMFB \cite{groth_prl17}) provides some improvement. Very good agreement, at low temperatures, was observed for DFT-MD simulations with PBE functionals \cite{Wang2013} confirming the crucial importance of correlation effects \cite{desjarlais_03}.

The presented comparisons are by no means exhaustive and do not pretend to be representative. The focus was on data that are available for the isotherms that were investigated in our paper (these values were selected due to existing RPIMC data), so additional errors, caused by interpolation, could be avoided. For these reasons we did not compare to other frequently used data sets including the SESAME tables  or the Rostock equation of state which was reported to be close to the RPIMC results \cite{Becker_2014}.
For more detailed comparisons, also with other methods, the reader is referred to our extensive data tables provided in the Appendix.

Due to the relative simplicity of hydrogen, first principle FP-PIMC simulations are possible that are free of systematic errors.
We expect that the benchmark data presented in our paper will allow one to constrain thermodynamic data to within $1\%$, providing ample opportunities to improve alternative simulation methods as well as chemical models for the challenging conditions of warm dense matter.  
This will be important not only for dense hydrogen but also  for the application of theoretical models and simulations to more complex materials and for reliable comparisons with existing and upcoming experiments.

%comparison to simple models, such as the Debye-Hückel model and chemical models, useful to understand the asymptotic behavior at low densities and high and low temperature, respectively.

  \section*{Acknowledgments}
   MB acknowledges stimulating discussions of the present results and their comparison to DFT simulations with P.R. Levashov and J. Vorberger. AF acknowledges discussions with V. Filinov on details of the FP-PIMC simulations.
   This work has been supported by the Deutsche Forschungsgemeinschaft via grant BO1366/15 and by the HLRN via grant shp00026.

\appendix

%%%%%%%%%%%%%%%%%%%%%

\section{Chemical Models (HSCM)}\label{ss:chem-model}
%
%\textcolor{red}{equation numbers only if they are cited in the text}\\
Here we briefly summarize the chemical models ``CM'' and ``HSCM'' that have been used for comparison in some of the figures of the main text.
We start with the grand potential and the ideal part of the free energy density 
\begin{eqnarray}
&&\Omega=F_{id}^{\text{HS}}+F_{id}^{C}+F^{\text{HS}}_{\text{ex}}+F^{C}_{\text{ex}}-\sum\limits_s \mu_s \rho_s,
\nonumber\\
&& \beta f_{id}=\sum_s \rho_s (\ln \Lambda_i^3 \rho_i-1) \,.
\nonumber
\end{eqnarray}
The free energy is decomposed into an ideal gas part, $F_{id}^{s}$, and an excess part, $F^s_{\text{ex}}$, related to  non-ideality effects. The contribution of  neutral particles (atoms/molecules, $s=\{a,m\}$), is treated on the hard sphere level (superscript ``HS''), whereas the superscript ``C'' denotes the Coulomb contribution ($s=\{e,i\}$). Charge-neutral contributions, e.g.~\cite{schlanges_cpp_95,Redmer} are neglected. 

Using the thermodynamic relations we can express
the pressure, the excess chemical potential and the excess interaction energy as follows
\begin{eqnarray}
&&p=\sum_s \mu_s \rho_s -(f^s_{id}+f^s_{ex}),
\nonumber\\
&&\mu^{a(m)}_{ex}=\frac{\partial f^{\text{HS}}_{ex}}{\partial \rho_{a(m)}},\quad \mu^{e(i)}_{ex}=\frac{\partial f^{\text{C}}_{ex}}{\partial \rho_{e(i)}}\label{mu_hs}
\nonumber\\
&& \frac{E_{ex}^{\rm HS(C)}}{V}=f_{ex}^{\rm HS(C)}-T\, \frac{\partial f_{ex}^{\rm HS(C)}}{\partial T}\,.
\label{EC}
%\nonumber
\end{eqnarray}

\subsection{Hard-sphere fluid model}\label{HS-model}
To obtain the chemical potential and the pressure for each species ``$s$'' in a multi-component hard-sphere model, we follow the method proposed by Hansen-Goos {\it et al.}~\cite{Goos}. They introduced an expansion of the related equation of state in terms of the powers of partial densities $\rho_s$ and the size of all components weighted by the density contributions~\cite{Rosenfeld}
\begin{eqnarray}
&&n_0=\sum\limits_s \rho_s,\quad
n_1=\sum\limits_s \rho_s R_s, \;\nonumber \\
&&n_2=\sum\limits_s \rho_s A_s, \quad
n_3=\sum\limits_s \rho_s V_s\label{exp_param}
\end{eqnarray}
where $R_s$, $A_s=4 \pi R_s^2$, and $V_s=4/3 \pi R_s^3$ are the radius, the surface area, and the volume of a sphere of species ``$s$''.
Systematic improvements of the original Carnahan-Starling equation of state (derived for a one-component system)
\begin{eqnarray}
\frac{\beta p}{\rho} =\frac{1+\eta+\eta^2-\eta^3}{(1-\eta)^3}   \,,
\nonumber
\end{eqnarray}
up to the third-order expansion in the packing fraction, $\eta = n_3$, have been analyzed in detail~\cite{Goos}. The validity of the resulting hard-sphere EOS has been justified by the accurate agreement with the simulation data. 

The third order expansion has been employed in the present analysis o dissociation equilibrium in hydrogen (deuterium), where atoms and molecules have been treated as spheres of radii $\{R_a,R_m\}$,  with the result for the pressure
\begin{eqnarray}\nonumber
\beta p^{(3)}=&&\frac{n_0}{1-n_3}+\frac{n_1 n_2 (1+1/3 n_3^2)}{(1-n_3)^2}+\\ \nonumber
&&\frac{n_2^3 (1-2/3 n_3+1/3 n_3^2)}{12\pi (1-n_3)^3}\,.
\end{eqnarray}
The excess chemical potentials (\ref{mu_hs}) follow as
\begin{eqnarray}
\mu_{ex}^s=\frac{\partial f^{(3)}_{ex}}{\partial n_3} \, V_s+\frac{\partial f^{(3)}_{ex}}{\partial n_2} \, A_s+\frac{\partial f^{(3)}_{ex}}{\partial n_1} \, R_s+\frac{\partial f^{(3)}_{ex}}{\partial n_0} \,, 
%\nonumber
\label{mu_atoms}
\end{eqnarray}
using the third-order $n_3$-expansion, $f^{(3)}_{ex}$~\cite{Goos},
for the excess free energy density. 

The chemical model that includes the hard sphere effects, as described above, has been called ``HSCM'', in the main text, whereas the model that neglects these terms has been denoted  ``CM''.
%The corresponding nonideality effect due to hard-spheres in the combined "HSEOS" is presented in Figs.~\ref{fig:PressEnergT95}-\ref{fig:EOST15}. 
The deviation from the classical ideal pressure and energy (excess contribution) can be quantified by
\begin{eqnarray}
 &&\beta \Delta p_{\text{HS}}=\beta p^{(3)}-n_0,
 %\label{dpHS}
 \nonumber\\   
 && \Delta E_{\text{HS}}=\frac{1}{n}\left[f^{(3)}_{ex}-T\, \frac{\partial f^{(3)}_{ex}}{\partial T}\right]. 
 \nonumber
% \label{dEHS}
\end{eqnarray}
The factor, $1/n$ (with $n=N/V$ being the total number density of ions (electrons), appears if we define $\Delta E_{\text{HS}}$ as the excess energy per ion, due to the ``HS''-effect, in a system consisting of both, neutrals and free particles. 
Note, that the contribution of the second term in $\Delta E_{\text{HS}}$ should be accurately evaluated, as the  fractions of atoms and molecules experience a noticeable variation with temperature and density. This results in a similar temperature-density dependence of the expansion variables~(\ref{exp_param}) constructed from corresponding partial densities of neutrals, $\rho_a(r_s,\beta)$ and $\rho_m(r_s,\beta)$. Their explicit evaluation can be obtained via the solution of the non-ideal Saha equations introduced in Sec.~\ref{a:saha}.

In the present model, we use fixed (temperature and density independent) effective radii $\{R_a,R_m\}$ choosing the  values $\{0.66\AA,1.32\AA \}$, obtained by the fluid variation theory of Juranek {\it al.}~\cite{FVT}. We also used other radii as provided by Ref.~\cite{schlanges_cpp_95}, but did observe significant changes in the considered parameter range, $r_s \gtrsim 3$.

\subsection{Non-ideal Saha equations for the ionization--dissociation equilibrium}\label{a:saha}
We impose electro-neutrality and charge conservation which leads
\begin{eqnarray}
 N_{i(e)}=N_{i(e)}^\star+N_a+2 N_m, \quad N_e=N_i=N\,,
\nonumber
\end{eqnarray}
with $N^{*}$ denoting the number of unbound electrons (ions), and $n=n_{i(e)}=N_{i(e)}/V$, the full ion (electron) density. 
The conditions for the dissociation-ionization equilibrium are given by
\begin{eqnarray}
\mu_m=\mu_a+\mu_a, \quad \mu_a=\mu_e+\mu_i\,,
\end{eqnarray}
where the chemical potential of the constituent particles are spitted into an ideal and an exchange-correlation contribution
\begin{eqnarray}
&& \mu^{id}_{e}=\mu^{cl}_{e}+\Delta \mu_e, \label{mu_id} 
\nonumber\\
&&\mu_e+\mu_i=\mu^{id}_{e}+\mu^{id}_{i}+\mu_{ex}^C,
\nonumber\\
&&\mu_{ex}^C=\frac{\partial f_{ex}^C}{\partial n_e}+\frac{\partial f_{ex}^C}{\partial n_i},\label{mu_C} 
\nonumber\\
&&\mu_{a}=\mu^{cl}_{a}+\mu_{ex}^a, \quad \mu_{m}=\mu^{cl}_{m}+\mu_{ex}^m.
\nonumber
\end{eqnarray}
$\mu^a_{ex}$ and $\mu^m_{ex}$ are defined in Eq.~(\ref{mu_atoms}) by the partial derivatives of the  free energy functional of a reference two-component hard-sphere system.
%in Eq.~(\ref{mu_atoms}) where we have defined the excess chemical potential for the ionization and dissociation process
The ideal part depends on the thermal wavelength $\lambda_s$, the spin degeneracy factor $g_s$, and the particle density $n_s$. For electrons and ions we take into account the ``excluded volume'' factor $(1-\eta)$ 
%(omitted for atoms/molecules)
\begin{eqnarray}
\beta \mu^{id}_{s}=\ln \left(\frac{n_s \lambda_s^3}{g_s(1-\eta)} \right).\label{mu_s} 
\end{eqnarray}
The term $\Delta \mu_e$ in Eq.~(\ref{mu_id}) accounts for the deviation of the ideal Fermi gas result from the classical expression~(\ref{mu_s}). For the free energy density  $f_{ex}^C$ one can employ a suitable approximation derived in
~\cite{blue-book,fortov-book}. In particular, we have used a quantum Debye-Hückel reduced-mass approximation where the free energy, the interaction energy and the chemical potential (due to Coulomb correlations) are expressed via Ebeling's  ``ring functions'', e.g. \cite{kremp1967,riemann1995,Ebeling2016}
\begin{eqnarray}
 &&\beta E_{ex}^{ij}/V=- \frac{\kappa^3}{8 \pi} G(\kappa \lambda_{ij}),\nonumber\\
 &&\mu_{ex}^{ij}=-\frac{1}{2} e^2 \kappa \, G(\kappa \lambda_{ij}),\nonumber\\
 &&\beta f_{ex}^{ij}=-\frac{\kappa^3}{12 \pi} R(\kappa \lambda_{ij})\,,\nonumber\\
 &&\beta p_{ex}^{ij}=-\frac{\kappa^3}{24 \pi}\left[G(\kappa \lambda_{ij})+\kappa \lambda_{ij} G^{\prime} (\kappa \lambda_{ij})\right]\,,
 \nonumber
\end{eqnarray}
%The $G$-function is obtained by the Debye charging procedure where the Coulomb potential was substituted with the two-body Kelbg potential, cf. the Kremp-Schmitz approximation~\cite{kremp1967,riemann1995,Ebeling2016}, 
which have been approximated by Pad\'{e} formulas~\cite{Ebeling_1982, Ebeling_1985, fortov-book}
\begin{eqnarray}
&& G(x)=\frac{1+0.1875 x^2+x^3}{1+0.4431 x+ 0.1963 x^2+1.1892 x^{7/2}},
\nonumber\\
&& R(x)=\frac{1+0.1875 x^2+3 x^3}{1+0.4431 x+ 0.1963 x^2+3.667 x^{7/2}}.
\nonumber
\end{eqnarray}
The argument, $x=\kappa \lambda_{ij}$, includes the inverse Debye length, $\kappa^2=8 \pi \bar n^{\star} \beta e^2$ (where $\bar n^{\star}$ is the re-normalized by the excluded volume effect density of free ions/electrons, see below), and $\lambda_{ij}^2=\hbar^2 \beta /2 \mu_{ij}$. In the  reduced mass approximation we assume that the differences between the atom and ion masses are not relevant and, therefore, we set in Eqs.~(\ref{EC}) and~(\ref{mu_C}) 
\begin{eqnarray}
 &&\mu^i_{ex}=\mu_{ex}^e=\mu_{ex}^{ei},\nonumber\\ 
 &&\mu_{ex}^C=\mu^i_{ex}+\mu_{ex}^e=2 \mu_{ex}^{ei},\nonumber \\
 &&E_{ex}^C=E_{ex}^{ei}, \quad p_{ex}^C=p_{ex}^{ei}.
 \nonumber
% \label{EP_C} 
\end{eqnarray}

Next, we consider the ionization equilibrium, 
%$a \rightleftarrows a^+ +e$, between the atoms and the ionized plasma component. This 
which leads to the Saha equation with the non-ideality effects included via the excess chemical potential $\mu_{ex}^I$ of the ionization process
\begin{eqnarray}
&&\frac{n_a}{n_{i}^\star n_{e}^\star}=\Lambda^3\, \sigma_{\text{PLB}}(\beta) \, \exp(\beta \mu_{ex}^I),\label{MAL1}\\
&&\mu_{ex}^I=\mu_{ex}^C+\Delta \mu_e - \mu^a_{ex}
\nonumber
\end{eqnarray}
where 
\begin{eqnarray}
&&\Lambda^2=\frac{\lambda_e^2 \lambda_i^2}{\lambda_a^2}= \frac{2 \pi \hbar^2 \beta}{m_r},
\nonumber
\\
&& \sigma_{\text{PLB}}(\beta)=\sum\limits_{s=1}^{\infty} s^2 \, \left[\exp(-\beta E_s)-1+\beta E_s\right]\,,
\nonumber
\end{eqnarray}
and $\sigma_{\text{PLB}}(\beta)$ is the regularized Planck-Brillouin-Larkin (PBL) partition function, for a discussion, see Refs.~\cite{PBL1,PBL2}.  
Now, the atom fraction can be determined from the Saha equation~(\ref{MAL1}) which, however, needs to be modified to take into account the excluded volume effect: 
electrons and ions ``cannot penetrate'' neutrals, i.e.
%It takes into account the Pauli blocking for free charged particles which cannot penetrate into interior of neutrals (atoms/molecules), cf. 
$V \rightarrow V(1-\eta)$. The 
 packing fraction,  $\eta=n_3$, see Eq.~(\ref{exp_param}), is estimated via the number density of atoms and molecules and their radii, and 
re-normalizes the free electron and ion density to $\bar{n}^\star=n^\star/(1-\eta)$. Introducing the ionization fraction, $n^\star=\alpha_I n$, the mass-action law~(\ref{MAL1}) can be reduced to
\begin{eqnarray}
 \frac{1-\alpha_I}{\alpha_I^2}= \frac{1}{(1-\eta)^2}\, K^I(\bar{n}^\star,\beta,\mu_{ex}^I), 
 \nonumber
 %\label{al_eq}
\end{eqnarray}
which can be solved with respect to $\alpha_I$ as a function of
the rate constant $K^I$ and the excess chemical potential 
\begin{eqnarray}
 &&K^I(\bar{n}^\star,\beta,\mu_{ex}^I)=\Lambda^3\, \sigma_{\text{PLB}}(\beta) \, \exp(\beta \mu_{ex}^I),
 \nonumber\\
 &&\mu^I_{ex}=f(\bar{n}^\star,\beta,\alpha_I,\alpha_D).
\nonumber
\end{eqnarray}
%Note, that both quantities experience a dependence on density, temperature and the relative fraction of neutrals defined by the ionization and the dissociation %coefficients $\{\alpha_I,\alpha_D\}$.

The second Saha equation for the dissociation equilibrium  can be written in the form
\begin{eqnarray}
&&\frac{n_m}{n_a n_a}=K^D(\beta,\mu_{ex}^D), \label{MAL2}\\
&&K^D(\beta,\mu_{ex}^D)=\frac{g_m}{g_a g_a}\frac{\Lambda_a^3 \Lambda_a^3}{\Lambda_m^3} Z^{\text{rot}} Z^{\text{vib}} \exp\left[\beta (D_0+\mu_{ex}^D)\right],\nonumber \\
&&\mu_{ex}^D=2 \mu^a_{ex}-\mu^m_{ex}
\nonumber
\end{eqnarray}
where $D_0=4.763$eV is the dissociation energy of a $H_2$-molecule, and the contributions of vibrational and rotational states~\cite{Mayer-book} are included via the partition functions
\begin{eqnarray}
&&Z^{\text{rot}}\approx \frac{T}{T_{\text{rot}}}\,, 
\nonumber
\\
&&Z^{\text{vib}}\approx \left[1-\exp(-\beta T_{\text{vib}})\right]^{-1}\,,
\nonumber
\end{eqnarray}
where we use $T_{\text{vib}}=87.58$K and $T_{\text{rot}}=6338.2$K. For temperatures $T>1000$K, in both expression, contributions of the order $\mathcal{O}(T_{\text{rot(vib)}}/T)$ in the free energy density can be neglected.  Introducing the degree of dissociation, $n_a=\alpha_D (n_a+n_m)$, the mass-action law~(\ref{MAL2}) becomes
\begin{eqnarray}
\frac{1-\alpha_D}{\alpha_D^2}=  K^D(\beta,\mu_{ex}^D). \label{al2_eq} 
\end{eqnarray}
%Finally, the Saha equations are completed by the particle number conservation %which inter-connect the ionization and dissociation factors
%\begin{eqnarray}
%n^\star_{i(e)}=n-n_a-2\, n_m. 
%\end{eqnarray}

%To summarize, the self-consistent solution of the coupled rate equations~(\ref{al_eq}),(\ref{al2_eq}) provide us with information on the plasma composition at different thermodynamic conditions.

Examples of solutions of the coupled Saha equations are included in Fig.~\ref{fig:Nfrac}. The comparison to the results of the FP-PIMC cluster analysis 
indicates remarkable agreement in a wide density range, $3\leq r_s\leq 100$,  cf.~Sec.~\ref{lowT} of the main text.

\section{Tables of FP-PIMC thermodynamic data}\label{eos-table}
In this appendix we present detailed tables of our FP-PIMC simulations. In addition to the equation of state and the  total energy we also present the relative deviation between our data and the RPIMC results of Militzer \textit{et al}, Ref.~\cite{Militzer_2021}. The data are converged with respect to the number $P$ of high-temperature factors and the particle number $N$. In some cases, results for several particle numbers are given to illustrate the magnitude of finite size effects. In particular, underlined numbers indicate deviations to RPIMC when we used $N=64$ particles.

\begin{center}
\begin{table*}[]
\caption{First principles FP-PIMC data for deuterium/hydrogen plasma, including pressure $p$ and internal energy $E/N$ with statistical errors given in the parantheses. The column $\Delta p/p$ denotes the relative statistical error (first number) and the deviation from RPIMC~\cite{Hu2011}, i.e. $(p^{\text{RPIMC}}-p^{\text{FP-PIMC}})/p^{\text{FP-PIMC}}$), second number, and similar for $\Delta E/E$. $\theta=T/T_F$ is the electron degeneracy parameter. The default system size is $N=34$. For $N=64$ (if available) an extra line is added.}
\label{tab:data}
\footnotesize
\begin{tabular}{  c c c c c c c c}
\hline
\hline
 $T$[K] & $r_s$ & $p$ [Mbar] & $E/N$ [eV] & $\Delta p/p\, [\%]$   & $\Delta E/E\, [\%]$ & $\theta$ \\
\hline

\multirow{24}{4em}{250,000} & 1.2 (N20) & 58.30(43) & 39.98(43) & 0.74(-0.85) & 1.08(-0.21) & 0.62  \\
 &1.4 (N20) & 35.616(32) & 40.69(5) & 0.09(0.12) & 0.13(0.75) & 0.84  \\
 &1.4 (N30) & 35.75(13)  & 40.97(20) & 0.35(-0.27) & 0.49(0.06) &   \\
&1.6 & 23.825(29) & 42.62(7) & 0.12(-0.40) & 0.17(-0.28) & 1.10  \\
&1.8 & 16.7343(87) & 44.19(3) & 0.052 & 0.067 & 1.39  \\
& 1.8 (N64)  & 16.737(54)  & 44.27(18) & 0.32 & 0.41 &   \\
&2.0 & 12.2728(40) & 45.852(19) & 0.033(-0.59) & 0.041(-0.49) & 1.72 \\
&    &  12.314(14)  & 46.08(6)  & 0.11(\underline{-0.93}) & 0.14(\underline{-1.03})&   \\

& 2.4  & 7.2001(16) &  48.699(13)  & 0.021(0.08) & 0.026(0.29)  & 2.47  \\
&       &  7.213(3)  & 48.86(3)     & 0.041(\underline{-0.096}) & 0.049(\underline{-0.041}) & \\

& 3.0  & 3.76445(62) & 52.087(9) & 0.017(-0.30) & 0.019(-0.21) & 3.87  \\
&      & 3.76812(92) & 52.202(15) & 0.024(\underline{-0.40}) & 0.028(\underline{-0.43}) & \\

& 4.0 & 1.62844(23) & 55.7648(87) & 0.014(-0.21) & 0.016(-0.1)  & 6.88  \\
&     & 1.63076(31) & 55.908(11)  & 0.019(\underline{-0.35})  & 0.020(\underline{-0.35})&\\

&5.0 & 0.84767(12) & 58.032(8) & 0.013(-0.19) & 0.015(-0.038)  & 10.75  \\
&    & 0.84859(15) & 58.16(1)  & 0.018(\underline{-0.31}) & 0.019(\underline{-0.26}) & \\

&6.0 &0.49589(7)  & 59.498(8) & 0.014(-0.26) & 0.015(-0.17)   & 15.47  \\
&    & 0.496403(95) & 59.6239(12) & 0.019(\underline{-0.36}) & 0.020(\underline{-0.37}) &  \\

&7.0 &0.31456(4)  & 60.485(9) & 0.014 & 0.014 & 21.06  \\
&    & 0.31478(6) & 60.592(12) & 0.019 & 0.02 &        \\

&8.0 &0.21186(3) &  61.199(9) & 0.014(-0.12) & 0.015(-0.03) & 21.06  \\
&    & 0.212146(40) & 61.343(12) & 0.019(\underline{-0.26}) & 0.02(\underline{-0.26}) &   \\

&10.0 &0.10923(2) & 62.130(9) & 0.014(-0.21) & 0.014(-0.06) &42.99  \\
&      & 0.109346(21) & 62.255(12) & 0.019(\underline{-0.32}) & 0.019(\underline{-0.26}) &  \\

&12.0 &0.063494(9) &  62.709(9) & 0.014(-0.15) & 0.014(-0.06) &61.90  \\
&     & 0.063537(12) & 62.807(12) & 0.019(\underline{-0.216}) & 0.019(\underline{-0.217}) & \\

&14.0 & 0.0401099(55) & 63.106(9) & 0.014(-0.17) & 0.014(-0.10)&84.26 \\
&     & 0.0401336(76) & 63.196(12) & 0.019(\underline{-0.23}) & 0.019(\underline{-0.25})&  \\

\hline
\multirow{40}{4em}{181,823} & 1.6 (N20) & 16.59(2) & 24.39(5)  & 0.12(0.42) & 0.19(1.48) &0.80  \\
& 1.6 & 16.82(16) & 24.98(37) &  0.93(-0.92) & 1.49(-0.94) & \\
&1.8 & 11.679(23) & 26.05(8) & 0.20 & 0.31 &1.01 \\
& 2.0 & 8.4965(64) & 27.225(30) & 0.075(-0.60) & 0.11(-0.53) &1.25 \\
& 2.2  & 6.4233(28) & 28.556(17) & 0.043(-0.42) & 0.061(-0.27)  &1.513 \\
&  2.2 (N64) & 6.447(14) & 28.747(84) & 0.21(\underline{-0.79}) & 0.29(\underline{-0.93}) &\\

&2.4 & 4.9830(15)  & 29.783(13) & 0.031(-0.08) & 0.042(0.20) & 1.800 \\
&    & 4.9906(48)  & 29.885(39) & 0.097(\underline{-0.23}) & 0.13(\underline{-0.15}) &  \\

&2.8 & 3.1889(7) & 32.002(9) & 0.022 & 0.028 & 2.45 \\
&    & 3.1936(14) &  32.103(18) & 0.043 & 0.055 &  \\

&3.0 & 2.6153(5) & 33.011(8) & 0.02(-0.35) & 0.025(-0.31) & 2.81 \\
&    & 2.6179(9) & 33.093(14) & 0.034(\underline{-0.46}) & 0.043(\underline{-0.55}) &  \\

&3.5 & 1.6781(3) & 35.1287(75) & 0.018 & 0.021 & 3.83 \\
&    & 1.67937(42) & 35.196(10) & 0.025 & 0.029 &  \\

&4.0 & 1.14179(18) & 36.812(7) & 0.016(-0.16) & 0.018(-0.032) & 5.00 \\
&    & 1.14300(24) & 36.898(9) & 0.021(\underline{-0.26}) & 0.025(\underline{-0.265}) &  \\

& 5.0 & 0.598717(85) & 39.283(6) & 0.014(-0.18) & 0.016(-0.083) & 7.82 \\
&     & 0.59918(12) & 39.3579(86) & 0.020(\underline{-0.26}) & 0.022(\underline{-0.274})   \\

&6.0 & 0.352316(55) & 40.955(7) & 0.016(-0.17) & 0.017(-0.086) & 11.25 \\
& & 0.352616(73) & 41.033(9) & 0.021(\underline{-0.26}) & 0.023(\underline{-0.275}) &  \\

&7.0 & 0.224470(34) & 42.115(7) & 0.015 & 0.016 &  15.32 \\
& & 0.224677(43) & 42.200(9) & 0.019 & 0.021&  \\

&8.0 & 0.151657(21) & 42.954(6) & 0.014(-0.10) & 0.015(-0.033) &  20.01 \\
& & 0.151794(29) & 43.0387(87) & 0.019(\underline{-0.19}) & 0.020(\underline{-0.23}) &   \\

&10.0 & 0.078559(11) & 44.0803(67) & 0.015(-0.19) & 0.015(-0.09) &  31.27 \\
&     & 0.078629(15) & 44.164(9)   & 0.019(\underline{-0.28}) & 0.020(\underline{-0.28}) &   \\

&12.0 & 0.0457687(63) & 44.748(6) & 0.014(-0.17) & 0.014(-0.06) &  45.02\\
&     & 0.0458167(86) & 44.8399(87) & 0.019(\underline{-0.28}) & 0.019(\underline{-0.27}) &  \\

&14.0& 0.0289573(40) & 45.2016(64) & 0.014(-0.13) & 0.014(-0.025) &  61.28 \\
&     & 0.0289812(55) & 45.2807(9) & 0.019(\underline{-0.21}) & 0.019(\underline{-0.20}) &  \\ %\cdashline{1-2}

&17.0& 0.0162524(22) & 45.656(6)   & 0.013 & 0.013 &  90.36 \\
&20.0&  0.0100110(13) & 45.9447(62) & 0.013 & 0.0135 & 125 \\
&25.0& 0.00514182(71) & 46.2451(64) & 0.014 & 0.014 & 195 \\
&30.0 & 0.00298112(42) & 46.4242(66) & 0.014 & 0.014 & 281 \\
&40.0 & 0.12607(18) $\cdot 10^{-3}$ & 46.6453(70) &  0.0145 & 0.015 & 500 \\
&50.0 & 0.646379(90) $\cdot 10^{-3}$ & 46.7662(66) & 0.014 & 0.014 &  782 \\
&60.0&  0.374232(53) $\cdot 10^{-3}$ & 46.8239(66) & 0.014 & 0.014 & 1125 \\
&80.0 & 0.158027(22)  $\cdot 10^{-3}$ & 46.9102(66) & 0.014 & 0.014 & 2001 \\
&100.0& 0.80937(11) $\cdot 10^{-4}$ & 46.9505(67) & 0.014 & 0.014  &  3127 \\

\hline

\end{tabular}
\end{table*}
\end{center}

%%%%%%%%%%%%%%%%%%%%

\begin{center}
\begin{table*}
%\caption{\label{Tab-EOS2}
%EOS table2
\begin{tabular}{  c c c c c c c c }
\hline
\hline
$T$[K] & $r_s$ & $p$ [Mbar] & $E/N$ [eV] & $\Delta p/p\, [\%]$   & $\Delta E/E\, [\%]$ & $\theta$ \\
\hline
\hline
\multirow{24}{4em}{125,000} & 1.6 (N20) & 10.85(17) & 9.87(40) & 1.58(1.06) & 4.11(5.78) &0.55  \\
&1.8 (N20) & 7.52(2) & 10.91(7)  &   0.27(-0.29) & 0.64(1.81) &0.69  \\
&2.0 (N20) & 5.4316(48) & 11.82(2) & 0.087(-0.34) & 0.19(0.74) &0.86\\
&2.2 (N20)&  4.0810(19) & 12.823(1) & 0.046(0.17) & 0.092(1.53) &1.04 &\\
&2.5 (N20) & 2.80676(81) & 14.3374(74) & 0.029(-0.505) & 0.052(1.42) &1.30  \\
&2.2 & 4.0941(74) & 12.937(46) & 0.18(-0.15) & 0.35(0.64) &1.04 \\
&2.5 & 2.8180(20) & 14.476(18) & 0.069(-0.90) & 0.12(0.45) &1.30  \\
&3.0 & 1.6652(5) & 16.7418(85) & 0.032(-0.55) & 0.051(-0.49) &1.93  \\
&3.5&  1.07448(24) & 18.747(6) & 0.023 & 0.032 & 2.63 \\
&4.0 & 0.73490(14) & 20.4006(54) & 0.019(-0.56) & 0.026(-0.54) &3.40  \\
&5.0 & 0.389230(62) & 22.966(5) & 0.016(-0.316) & 0.020(-0.246) &5.38 \\
&6.0 & 0.230966(36) & 24.8116(46) & 0.016(-0.20) & 0.018 (-0.087) &7.73 \\
&7.0 & 0.148395(23) & 26.2034(46) & 0.015 & 0.0176 &10.53 \\
&8.0 & 0.100851(14) & 27.2199(43) & 0.014(-0.050) & 0.016(0.00036) & 13.75 \\
&10.0 & 0.052688(8) & 28.6154(47) & 0.015(-0.091) & 0.0165(0.0160) & 21.5 \\
&12.0& 0.0308841(43) & 29.489(4) & 0.0138(-0.142) & 0.0147(0.00237) & 31.0  \\
&14.0 & 0.0196128(27) & 30.067(4) & 0.0136(-0.0652) & 0.0144(0.0422) & 42.1 \\
%\cdashline{1-2}
&17.0 & 0.0110505(15) & 30.6457(42) & 0.0134 & 0.0138 &  62.12 \\
&20.0 & 0.682187(91) $\cdot 10^{-2}$ & 31.0026(4) & 0.0133 & 0.0136 & 86.0 \\
&25.0  & 0.351243(47) $\cdot 10^{-2}$ & 31.376(4) & 0.0134 & 0.0136 & 134.0 \\
&30.0 & 0.204159(30) $\cdot 10^{-2}$ & 31.632(5)& 0.0145 & 0.0147 & 193  \\
&40.0 & 0.86397(12) $\cdot 10^{-3}$ & 31.8596(44)& 0.0138 & 0.014 & 343  \\
&50.0 & 0.443236(59) $\cdot 10^{-3}$ & 31.9919(43)& 0.0133 & 0.0134 & 537 \\
&60.0 & 0.256795(34) $\cdot 10^{-3}$ & 32.0704(43)&0.0133 & 0.0134 & 773  \\
&80.0 & 0.108558(15) $\cdot 10^{-3}$ & 32.1847(44) & 0.0137 & 0.0137 & 1375 \\
&100.0& 0.556326(82) $\cdot 10^{-4}$ &  32.2407(48) & 0.0148 & 0.0148 & 2149  \\

\hline
\multirow{30}{4em}{95,250} & 2.2 & 2.905(33) & 4.84(20)  &  1.135(1.69) & 4.22(9.59) & 0.79 \\
&2.5 & 2.0060(42) & 6.243(38) & 0.208 & 0.61 & 1.02 \\
&3.0 & 1.17896(67) & 8.109(10) & 0.057(-0.25) & 0.13(0.38) & 1.47 \\
&3.0 (N64) & 1.1856(25) & 8.230(40) & 0.21(\underline{-0.81}) & 0.48(\underline{-1.09}) & 1.47 \\

&4.0 & 0.52179(12) & 11.3647(47) & 0.024(-0.34) & 0.042(0.046)& 2.62  \\
& & 0.52180(23) & 11.3797(88) & 0.045(\underline{-0.345}) & 0.078(\underline{-0.085}) & 2.62 \\

&5.0 & 0.278393(52) & 13.851(4) & 0.019(-0.14) & 0.027(-0.22) & 4.09 \\
& & 0.278450(84) & 13.8713(62) & 0.030(\underline{-0.16})& 0.045(\underline{-0.37})  & 4.09 \\

&6.0 & 0.166672(27) & 15.7828(34) & 0.016(-0.34) & 0.022(-0.27) & 5.90 \\
& & 0.166827(36) & 15.8144(46) & 0.0216(\underline{-0.435}) & 0.029(\underline{-0.47}) & 5.90 \\

&7.0 & 0.107805(16) & 17.2732(33) & 0.015 & 0.019 & 8.02 \\
& & 0.107891(22) & 17.3055(45) & 0.020 & 0.026  & 8.02 \\

&8.0 & 0.073792(11) & 18.4462(33) & 0.015(-0.11) & 0.018(0.02) & 10.48 \\
& & 0.073826(15) & 18.4723(44) & 0.0197(\underline{-0.16}) & 0.0238(\underline{-0.12}) & 10.48 \\

&10.0 & 0.0389322(58) &  20.0895(35) & 0.015(-0.16) & 0.017(-0.048) & 16.4  \\
& & 0.0389756(80) &  20.1351(47) & 0.020(\underline{-0.27}) & 0.0236(\underline{-0.274})& 16.4 \\

&12.0 & 0.0229801(33) & 21.1608(34) & 0.014(-0.044) & 0.016(0.091)& 23.6  \\
& & 0.0230113(53) & 21.2114(64) & 0.023(\underline{-0.18}) & 0.0256(\underline{-0.148})& 23.6 \\

&14.0& 0.0146659(21)& 21.8805(34) & 0.014(-0.176) & 0.016(-0.048) & 32.1\\
& & 0.0146822(30) & 21.9259(48) & 0.020(\underline{-0.287}) & 0.022(\underline{-0.255})& 32.1 \\
%\cdashline{1-2}
&17.0& 0.83024(12)  $\cdot 10^{-2}$ & 22.5957(33) &0.014  & 0.015 & 47.3 \\
& & 0.83092(16) $\cdot 10^{-2}$ & 22.6365(46) & 0.019 & 0.020 & 47.3  \\

&20.0 & 0.514349(71) $\cdot 10^{-2}$ & 23.0538(33) & 0.0137 & 0.014 & 65.5 \\
&25.0 & 0.265675(37) $\cdot 10^{-2}$ & 23.5141(34) & 0.014 & 0.0143 &  102.4  \\
&30.0 & 0.155475(32) $\cdot 10^{-2}$ & 24.0883(50) & 0.0205 & 0.021 &  147.4  \\
&40.0 & 0.655798(95) $\cdot 10^{-3}$ & 24.0844(35) & 0.014 & 0.0146 & 262.0 \\
&50.0 & 0.336680(50) $\cdot 10^{-3}$ & 24.2311(36) & 0.0148 & 0.0149 & 409.0  \\
&60.0& 0.195317(27) $\cdot 10^{-3}$ & 24.3396(34) & 0.014 & 0.014 &  589.6 \\
&80.0 &0.825642(12)  $\cdot 10^{-4}$ & 24.443(3) & 0.014 & 0.014 & 1048 \\
&&  0.826306(94) $\cdot 10^{-4}$ & 24.4754(28)& 0.0114 & 0.0115 & 1048  \\
&100.0 & 0.423114(60) $\cdot 10^{-4}$ & 24.4946(34) & 0.014 & 0.014 & 1638 \\
\end{tabular}
\end{table*}
\end{center}

%%%%%%%%%%%%%%%%%%%%

\begin{center}
\begin{table*}
%\caption{\label{Tab-EOS2}
%EOS table2
\begin{tabular}{  c c c c c c c c c}
%\hline
\hline
$T$[K] & $r_s$ & $p$ [Mbar] & $E/N$ [eV] & $\Delta p/p\, [\%]$   & $\Delta E/E\, [\%]$ & $\theta$  \\
\hline
\hline
\multirow{9}{4em}{75,000} & 3.0  & 0.8659(48) & 2.49(8) &0.55 & 3.1 & 1.16 \\
&4.0 & 0.38091(30)  & 5.098(11)  & 0.08 & 0.22 & 2.06 \\
&5.0 & 0.204275(94) & 7.3549(70) & 0.046 & 0.094 & 3.22  \\
&6.0 & 0.122951(45) & 9.1967(58) & 0.037 & 0.063 & 4.64  \\
&7.0 & 0.080013(32) & 10.7006(64) & 0.040 & 0.055 & 6.32 \\
&8.0 & 0.055101(18) & 11.9415(56) & 0.034  & 0.047 & 8.25  \\
&10.0 & 0.0294190(88) & 13.816(6) & 0.030 & 0.038 &  12.9  \\
&12.0 & 0.017534(54) & 15.1083(56) & 0.031 & 0.037 & 18.6  \\
&14.0 & 0.0112658(37) & 16.0037(62) & 0.033 & 0.039 & 25.28 \\
\hline

\hline
\multirow{30}{4em}{62,500} & 2.6 (N20) & 1.044(12) & -2.10(12) & 1.17(0.37) & 5.9(-9.5)  &0.726 \\

& 3.0  & 0.6901(71)  &  -0.95(11)  &1.03(-2.48) & 11.7(17.1)  &0.967 \\
&3.5& 0.4343(15) & 0.069(37) & 0.34 & 54. &1.32 \\
&3.5 (N64) & 0.393(22)  & -0.93(55) & 5.62(-11.) & 59.2(-179.) & 1.32 \\

&4.0 & 0.29689(41) & 1.1610(16) & 0.14(-0.64) & 1.32(0.73) & 1.70\\
&& 0.2966(13)  & 1.1579(50) & 0.44(\underline{-0.55}) & 4.25(\underline{1.04}) & 1.70 \\

&4.5 & 0.21366(18) & 2.194(9) & 0.082 & 0.43 & 2.17 \\
& & 0.21529(62) & 2.290(33) & 0.29 & 1.44 & 2.17 \\

&5.0 & 0.1592(1) & 3.137(7) & 0.063(-0.77) & 0.23(-1.60) & 2.70 \\
&    & 0.15961(18) & 3.167(13) & 0.11(\underline{-1.0}) & 0.41(\underline{-2.53})& 2.70\\

&6.0 & 0.096076(43) & 4.815(5)  & 0.04(-0.60) & 0.11(-0.88) &3.86 \\
&    & 0.096196(53) &  4.8309(68) & 0.055(\underline{-0.72}) & 0.14(\underline{-1.20}) & 3.86\\

&7.0 & 0.062783(24) & 6.2589(48) & 0.037 & 0.076  &5.27 \\
& & 0.062850(36) & 6.2774(72) & 0.057 & 0.12 & 5.27\\

&8.0 & 0.043290(15) & 7.4536(46) & 0.035(-0.09) & 0.062(0.31) & 6.87 \\
& & 0.043430(22)  & 7.5013(66) & 0.05(\underline{-0.415}) & 0.088(\underline{-0.32}) & 6.87 \\

&10.0 & 0.0233774(71)  & 9.450(4) & 0.03(-0.54) & 0.045(-0.52) & 10.75 \\
& & 0.02337(1)  & 9.4544(61) & 0.044(\underline{-0.525}) & 0.064(\underline{-0.56}) & 10.75 \\

&12.0 & 0.0140376(41) & 10.8839(43) & 0.029(-0.41) & 0.040(-0.22) & 15.5 \\
& & 0.0140452(60) & 10.8992(62) & 0.043(\underline{-0.46}) & 0.057(\underline{-0.36}) & 15.5 \\

&14.0 & 0.90889(27) $\cdot 10^{-2}$ & 11.9542(45) & 0.029(-0.47) &  0.038(-0.20) & 21.0 \\
& & 0.90993(40) $\cdot 10^{-2}$ & 11.9828(67) & 0.044(\underline{-0.58}) & 0.056(\underline{-0.44}) & 21.0 \\

%\cdashline{1-2}
&17.0 & 0.52209(15)  $\cdot 10^{-2}$ & 13.0526(46) & 0.029 & 0.035 & 31.0 \\
&20.0&  0.3267(1) $\cdot 10^{-2}$ & 13.7865(48) & 0.030 & 0.035 & 43.0 \\

&25.0 & 0.170665(61) $\cdot 10^{-2}$ & 14.5465(56) & 0.036 & 0.039 & 67.0 \\
& & 0.17071(8) $\cdot 10^{-2}$ & 14.5601)(73) & 0.047 & 0.050 & 67.0 \\

&30.0 & 0.99795(41) $\cdot 10^{-3}$  & 14.9682(65) & 0.041 & 0.043 & 96.5 \\

&40.0 & 0.42666(15) $\cdot 10^{-3}$  & 15.4458(56) & 0.035 & 0.036 & 171 \\
& & 0.42736)(19) $\cdot 10^{-3}$ & 15.4848(79) & 0.044 & 0.045 & 171  \\

&50.0 & 0.220030(68) $\cdot 10^{-3}$ & 15.6877(50) & 0.031 & 0.031 & 268.7 \\
& & 0.2202(1) $\cdot 10^{-3}$ & 15.7203(72) & 0.045 & 0.046 & 268.7  \\

&60.0 & 0.127664(38) $\cdot 10^{-3}$ & 15.8019(48) & 0.030 & 0.030 & 386 \\
& & 0.127874(57) $\cdot 10^{-3}$ & 15.842(7) & 0.044 & 0.045 & 386\\

&80.0 & 0.54130(16) $\cdot 10^{-4}$ & 15.9597(48) & 0.030 & 0.030 & 687 \\
&100.0 & 0.277896(80) $\cdot 10^{-4}$ & 16.0419(47) &0.029 & 0.029 & 1074 \\

\hline
\multirow{17}{4em}{50,000} & 3.0  & 0.5005(41)  & -4.556(64)  &0.82 & 1.41 & 0.77 \\
&4.0 &0.21794(81) & -2.749(30) & 0.37 & 1.10 & 1.38  \\
&5.0 & 0.11642(13) & -1.1855(95) & 0.11 & 0.80 & 2.15  \\
&6.0 & 0.070107(42) & 0.1824(54) & 0.060 & 2.95 & 3.09  \\
&7.0 & 0.045865(22) & 1.4186(44) & 0.046 & 0.31 & 4.21  \\
&8.0 & 0.031773(13) & 2.5272(41) & 0.042 & 0.16 & 5.50  \\
&10.0 & 0.0172604(60) & 4.4277(37) & 0.035 & 0.083 & 8.60  \\
&12.0 & 0.0104598(49) & 5.9609(54) & 0.047 & 0.091 & 12.40  \\
&14.0 & 0.68375(24) $\cdot 10^{-2}$ & 7.1964(43) & 0.035 & 0.06 & 16.85  \\
%\cdashline{1-2}
&17.0 & 0.39738(12) $\cdot 10^{-2}$ & 8.5612(37) &0.030 & 0.044 &  24.85 \\
&20.0 & 0.251274(71) $\cdot 10^{-2}$ & 9.5452(36) & 0.028 & 0.038 & 34.40 \\
&25.0 & 0.132898(42) $\cdot 10^{-2}$ & 10.6277(40) & 0.031& 0.038 & 53.74  \\
&30.0 & 0.78366(31) $\cdot 10^{-3}$ & 11.2675(50) & 0.039 & 0.044 & 77.4  \\
&40.0 & 0.33759(14) $\cdot 10^{-3}$ & 11.9635(53) & 0.042 & 0.044 & 137.5  \\
&50.0 & 0.174537(78) $\cdot 10^{-3}$ & 12.2841(57) & 0.045 & 0.046 & 215  \\
&60.0 & 0.101562(55) $\cdot 10^{-3}$ & 12.4629(70) & 0.054 & 0.056 & 309  \\
&100.0&  0.22157(12)  $\cdot 10^{-4}$ & 12.7506(69) & 0.053 & 0.054 & 860 \\

\end{tabular}
\end{table*}
\end{center}

%%%%%%%%%%%%

\begin{center}
\begin{table*}
%\caption{\label{Tab-EOS4}
%EOS table4
\begin{tabular}{  c c c c c c c c c}
%\hline
\hline
$T$[K] & $r_s$ & $p$ [Mbar] & $E/N$ [eV] & $\Delta p/p\, [\%]$   & $\Delta E/E\, [\%]$ & $\theta$  \\
\hline
\hline
\multirow{18}{4em}{31,250} & 4.0 (N28)  & 0.1158(31) & -8.25(11) & 2.70(-0.68) & 1.40(-1.42) & 0.86 \\
&5.0 & 0.0610(4) & -7.451(30) & 0.65(-1.9)  & 0.40(-1.11)  & 1.34 \\
&6.0 & 0.03624(23)& -6.928(29) & 0.63(-1.87) & 0.41(0.28) & 1.93 \\
&7.0 & 0.023019(79) & -6.454(16) & 0.34  & 0.247 & 2.63 \\
&8.0& 0.015877(42) & -5.892(13) & 0.26(-0.48) & 0.21(-0.55) &3.44 \\
&10.0 & 0.85724(71) $\cdot 10^{-2}$&  -4.8223(42) & 0.083(-1.43)  & 0.088(0.39) & 5.37 \\
&12.0 & 0.51871(36) $\cdot 10^{-2}$ & -3.8193(37) & 0.069(-1.58) &  0.098(0.019) & 7.74 \\
&14.0 & 0.33920(22) $\cdot 10^{-2}$ & -2.8807(37) & 0.065(-0.82) & 0.13(-0.47)  & 10.53 \\
&17.0 & 0.19977(11) $\cdot 10^{-2}$ & -1.5625(34) & 0.055 & 0.22 & 15.5  \\
&20.0 & 0.128679(67) $\cdot 10^{-2}$ & -0.3531(35) & 0.052 & 0.99 & 21.5 \\
&25.0 & 0.70367(34) $\cdot 10^{-3}$ & 1.3719(36) & 0.048 & 0.26 & 33.6 \\
&30.0 & 0.42727(19)$\cdot 10^{-3}$ & 2.7425(35) & 0.045 & 0.13 & 48.4 \\
&40.0 & 0.1934(1) $\cdot 10^{-3}$  & 4.7058(41) & 0.051 & 0.086 & 86. \\
&50.0 & 0.103092(56) $\cdot 10^{-3}$ & 5.86616  & 0.054 & 0.073  & 134.4 \\
&60.0 & 0.610953(35) $\cdot 10^{-4}$ & 6.5628(46) &0.058 & 0.067 & 344. \\
&100.0 & 0.136866(93) $\cdot 10^{-4}$ & 7.6035(54) & 0.068 & 0.072 & 536. \\

%\end{tabular}
%\end{table*}
%\end{center}

%\begin{center}
%\begin{table*}
%\caption{\label{Tab-EOS3}
%EOS table3
%\begin{tabular}{  c c c c c c c c c}
%\hline
%\hline
%$T$[K] & $r_s$ & $p$ [Mbar] & $E/N$ [eV] & %$\Delta p/p\, [\%]$   & $\Delta E/E\, [\%]$ & %$\theta$  \\
%\hline
\hline
\multirow{18}{4em}{15,625} & 4.0\footnote{{\rm extrapolation using HSCM, cf. Fig.~\ref{fig:convT15rs4}}}  & 0.0452(13) & -12.319(47) &2.84(6.0) & 0.38(-0.89) &0.43 \\
&$4.0$(N14) & 0.0430(13) & -12.47(5) & 3.(11.) & 0.4(-2.0) &0.43 \\
&5.0 & 0.02409(52) & -12.057(38) & 2.17(5.3) & 0.32(-0.72) &0.67 \\
&6.0 & 0.01484(14) & -11.846(18) & 0.97(0.46) &  0.15(-0.22) &0.967 \\
&7.0 & 0.974(14) $\cdot 10^{-2}$ & -11.669(28) & 1.44 & 0.24 & 1.32 \\
&8.0 & 0.6743(50) $\cdot 10^{-2}$ & -11.535(15) & 0.74(-2.27) & 0.13(0.126) &1.72 \\
&10.0 & 0.3572(27) $\cdot 10^{-2}$ & -11.370(51) & 0.75(-2.29) &  0.44(-0.79) &2.69 \\
&12.0 & 0.21171(98) $\cdot 10^{-2}$ & -11.2975(98) & 0.465(-3.26) &  0.087(-2.90) &3.87 \\
&14.0 & 0.13437(54) $\cdot 10^{-2}$ & -11.2028(86) & 0.40(-4.0) & 0.077(-3.33)&5.27 \\
&17.0 & 0.7617(18) $\cdot 10^{-3}$  & -11.0754(5) & 0.25  & 0.048&7.76 \\
&20.0 & 0.4685(11) $\cdot 10^{-3}$ & -10.9734(5) & 0.23  & 0.045 &10.75 \\
&25.0 & 0.24568(53) $\cdot 10^{-3}$ & -10.724(5) & 0.22 &  0.045 &16.8 \\
&30.0 & 0.14335(27) $\cdot 10^{-3}$  & -10.526(4) & 0.19 & 0.042  &24.2 \\
&40.0 & 0.6195(12) $\cdot 10^{-4}$ & -10.062(5)  & 0.20 & 0.048  &43.0 \\
&50.0 & 0.32544(58) $\cdot 10^{-4}$ & -9.527(4) & 0.18 & 0.046 & 67.2 \\
&60.0 & 0.19336(35) $\cdot 10^{-4}$ & -8.9906(45) & 0.18 & 0.05 & 96.7 \\
&80.0 & 0.8611(11) $\cdot 10^{-5}$  & -7.8957(36) &0.13 & 0.045 & 172. \\
&100.0 & 0.46378(58) $\cdot 10^{-5}$ & -6.7590(36) & 0.125 & 0.053 & 268. \\

\end{tabular}
\end{table*}
\end{center}

%%%%%%%%%%%%%%%%%%%%

%merlin.mbs apsrev4-1.bst 2010-07-25 4.21a (PWD, AO, DPC) hacked
%Control: key (0)
%Control: author (0) dotless jnrlst
%Control: editor formatted (1) identically to author
%Control: production of article title (0) allowed
%Control: page (1) range
%Control: year (0) verbatim
%Control: production of eprint (0) enabled
%

%%%%%%%%%%%%%%%%%%%%
%\newpage
%\bibliography{ref,mb-ref,al-ref}
%\bibliography{mybib}{}
%\bibliographystyle{plain}
% \include{dusty_manus.bbl}

% \begin{thebibliography}{10}
% \end{thebibliography}
\end{document}